\documentclass[aps,prb,twocolumn,superscriptaddress,showkeys]{revtex4-1}
\usepackage{graphicx}
\usepackage{subfigure}
\usepackage{appendix}
\usepackage{amssymb}
\usepackage{amsmath}
\usepackage[normalem]{ulem}
\usepackage{color}

\newcommand{\BFA}{BaFe$_2$As$_2$}

\newcommand{\ket}[1]{\left | #1 \right \rangle }

\newcommand{\braket}[2]{\left \langle #1 \right | \left . #2 \right \rangle }
\newcommand{\bracket}[3]{\left \langle #1 \right | #2 \left | #3 \right \rangle }
\begin{document}

\title{Description of resonant inelastic x-ray scattering in correlated metals}

\author{Keith Gilmore}
\thanks{These authors contributed equally}
\affiliation{Condensed Matter Physics and Materials Science Department, Brookhaven National Laboratory, Upton, NY 11973, USA}
\email[email: ]{kgilmore@bnl.gov}
\author{Jonathan Pelliciari}
\thanks{These authors contributed equally}
%\email[email: ]{pelliciari@bnl.gov}
\affiliation{Photon Science Division, Paul Scherrer Institut, CH-5232 Villigen PSI, Switzerland}
\affiliation{Department of Physics, Massachusetts Institute of Technology, Cambridge, MA 02139, USA}
\affiliation{National Synchrotron Light Source-II, Brookhaven National Laboratory, Upton, NY 11973, USA}
\email[email: ]{pelliciari@bnl.gov}
\author{Yaobo Huang}
\affiliation{Photon Science Division, Paul Scherrer Institut, CH-5232 Villigen PSI, Switzerland}
\affiliation{Beijing National Lab for Condensed Matter Physics, Institute of Physics, Chinese Academy of Sciences, Beijing 100190, China}
\author{Joshua J. Kas}
\affiliation{Department of Physics, University of Washington, Seattle, WA 98195, USA}
%\email[email: ]{jjkas@uw.edu}
\author{Marcus Dantz}
\affiliation{Photon Science Division, Paul Scherrer Institut, CH-5232 Villigen PSI, Switzerland}
\author{Vladimir N. Strocov}
\affiliation{Photon Science Division, Paul Scherrer Institut, CH-5232 Villigen PSI, Switzerland}
\author{Shigeru Kasahara}
\affiliation{Department of Physics, Kyoto University, Sakyo-ku, Kyoto 606-8502, Japan}
\author{Yuji Matsuda}
\affiliation{Department of Physics, Kyoto University, Sakyo-ku, Kyoto 606-8502, Japan}
\author{Tanmoy Das}
\affiliation{Department of Physics, Indian Institute of Science, Bangalore 560012, India}
\author{Takasada Shibauchi}
\affiliation{Department of Advanced Materials Science, University of Tokyo, Kashiwa, Chiba 277-8561, Japan}
\author{Thorsten Schmitt}
\email[email: ]{thorsten.schmitt@psi.ch}
\affiliation{Photon Science Division, Paul Scherrer Institut, CH-5232 Villigen PSI, Switzerland}

\date{\today}

\begin{abstract}
To fully capitalize on the potential and versatility of resonant inelastic x-ray scattering (RIXS), it is essential to develop the capability to interpret different RIXS contributions through calculations, including the dependence on momentum transfer, from first-principles for correlated materials. Toward that objective, we present new methodology for calculating the full RIXS response of a correlated metal in an unbiased fashion.  Through comparison of measurements and calculations that tune the incident photon energy over a wide portion of the Fe L$_3$ absorption resonance of the example material BaFe$_2$As$_2$, we show that the RIXS response in \BFA~is dominated by the direct channel contribution, including the Raman-like response below threshold, which we explain as a consequence of the finite core-hole lifetime broadening.  Calculations are initially performed within the first-principles Bethe-Salpeter framework, which we then significantly improve by convolution with an effective spectral function for the intermediate-state excitation.  We construct this spectral function, also from first-principles, by employing the cumulant expansion of the Green's function and performing a real-time time dependent density functional theory calculation of the response of the electronic system to the perturbation of the intermediate-state excitation.  Importantly, this allows us to evaluate the indirect RIXS response from first-principles, accounting for the full periodicity of the crystal structure and with dependence on the momentum transfer.

\end{abstract}

\maketitle

\section{Introduction} 

Resonant inelastic x-ray scattering (RIXS) is a highly versatile and powerful probe of elementary excitations in materials owing to its sensitivity to all electronic ({\em i.e.} spin, charge, and orbital) and lattice degrees of freedom, elemental and orbital selectivity, detailed polarization analysis, and ability to detect small sample volumes \cite{ament_resonant_2011,dean_insights_2015}. RIXS has been used to detect local and collective excitations such as \textit{dd} excitations \cite{ellis_correlation_2015,dantz_quenched_2016,fatuzzo_spin-orbit-induced_2015,das_spin-orbital_2018,fabbris_doping_2017,lu_dispersive_2018}, charge transfer \cite{lee_charge-orbital-lattice_2014}, phonons \cite{geondzhian_demonstration_2018,chaix_dispersive_2017,devereaux_directly_2016}, spin excitations \cite{ishii_observation_2017,dean_high-energy_2013,dean_itinerant_2014,pelliciari_magnetic_2016,zhou_persistent_2013,braicovich_dispersion_2009,ishii_high-energy_2014,lee_asymmetry_2014,peng_magnetic_2015,minola_collective_2015,dean_spin_2012,le_tacon_intense_2011,pelliciari_local_2017,chaix_dispersive_2017,dellea_spin_2017}, and other quasi-particles \cite{schlappa_spin-orbital_2012,monney_mapping_2012,hepting_three-dimensional_2018,chaix_dispersive_2017}. This has allowed the study of the key interactions characterizing materials such as electron-phonon coupling, magnetic exchange, fractionalization, and the interplay of emergent collective excitations.  Recently, the technique has even been extended to time-resolved experiments \cite{dean_ultrafast_2016,wang_theoretical_2018,Parchenko_V2O3_PRR}. The success of RIXS is due to the huge advancements in x-ray instrumentation \cite{strocov_high-resolution_2010,ghiringhelli_saxes_2006} and the concomitant development of RIXS theory, which has guided the interpretation of experimental data from studies of the cross section and the coupling between the RIXS signal and the corresponding excitations \cite{ament_resonant_2011,jia_persistent_2014,jia_using_2016,benjamin_probing_2015,kanasz-nagy_resonant_2015,lu_nonperturbative_2017,geondzhian_demonstration_2018,geondzhian_polarons_2020}.

Experimental RIXS efforts have focused to a large extent on strongly correlated electron materials, particularly Mott insulators.  Correspondingly, computational efforts typically utilize either model Hamiltonians on 1D or 2D lattices \cite{Maekawa_PhysRevLett.91.117001,Kotani_JPSJ.75.044702,Sawatzky_PhysRevB.77.104519,Daghofer_PhysRevB.85.064423,Johnston_SciRep.8.11080} or treat small clusters around the absorbing element at various approximations \cite{Josefsson_JPhysChemLett.3.3565,Haverkort_JPhysConfSer.712.012001,Hariki_PhysRevB.101.115130,Maganas_JPhysChemC.35.20163}.  The computational demands of both approaches scale prohibitively with system size, requiring truncation both with respect to the lattice or cluster size and with respect to the number of bands or active orbitals included in the calculation.  These restrictions are tolerable for many of the transition metal oxides commonly probed experimentally thus far, such as the cuprates and iridates, because of their insulating or shallow band character, the nearly complete $d$ shell or subshell and the typical reduced dimensionality of these materials.  However, these computational limitations appear to preclude accurate treatment of more itinerant systems or even multi-orbital non-metallic materials with moderate bandwidth.

Multi-orbital correlated materials present a rich physical landscape of interactions leading to a plethora of intriguing phenomena including spin, charge, and orbital orders, giant- and colossal-magneto resistance, metal-to-insulator transitions, and multiferroicity \cite{deng_natcomm, stadler_hund-mott,catalan_progress_2008,catalano_rare-earth_2018,fiebig_evolution_2016,hwang_emergent_2012}.  Correlated metals offer several examples of unconventional superconductivity distinct from cuprates including intermetallic heavy fermion systems \cite{Pfleiderer_RevModPhys.81.1551}, the iron-based superconductors \cite{stewart_rmp}, and Sr$_2$RuO$_4$ \cite{Maeno_Nature.372.532}.
Meanwhile, the narrow gap semiconductor FeSb$_2$ \cite{bentien_FeSb2-seebeck} and ultrathin films of FeSe \cite{shimizu_thermoelectric} possess exceptional thermoelectric properties.  These multiorbital systems are subject not only to charge correlations, characterized by the Hubbard interaction, but also orbital correlations determined by Hund's coupling.  The latter interaction leads to orbital differentiation and the possibility of orbital selective Mott transitions \cite{song_mott_2016}.  The orbital selective nature of correlations within Hund's metals and their interplay with Mott physics remains a fascinating and vigorously investigated subject \cite{Deng_NatComm.10.2721,Stadler_AnnalsPhysics_2019}.  While these properties have been probed by other techniques such as angle-resolved photoemission \cite{matt_$textbackslashmathrmnafe_0.56textbackslashmathrmcu_0.44textbackslashmathrmas$_2016, Zabolotnyy_Nature_2009}, optical conductivity \cite{Charnukha_JPCM_2014} and inelastic neutron scattering \cite{song_mott_2016, Liu_NatPhys_2012}, applying RIXS to correlated metals offers several advantages such as bulk, elemental and orbital sensitivity. However the development of RIXS on these materials is hampered by the lack of a theoretical framework to interpret spectra of metallic and multiorbital systems, limiting the utility of RIXS.

Experimentally, metals, and especially correlated metals, have only been marginally investigated in terms of their RIXS response and cross section due to their large fluorescence backgrounds that tend to obscure the more interesting contributions from collective excitations such as phonons, magnons and secondary charge excitations.  Magnon and paramagnon excitations have been studied by RIXS in parent and doped \BFA~compounds \cite{zhou_persistent_2013, pelliciari_reciprocity_2019, pelliciari_local_2017,garcia_anisotropic_2019}, in 111 \cite{pelliciari_intralayer_2016} and 1111  \cite{pelliciari_presence_2016} systems, and in FeSe \cite{rahn_paramagnon_2019,pelliciari_evolution_2020}.  However, the interpretation of RIXS data in multi-orbital systems remains challenging and to analyze the experimental spectra one must presently resort to model descriptions to subtract the fluorescence contribution and uncover the signatures of collective spin and lattice modes.  Improving upon this situation requires theoretical and computational advancements.

Here, we map the RIXS loss profile of \BFA~as a function of the incident photon energy across the Fe L$_3$ edge.  The most prominent feature in the experimental data is the presence of a Raman-to-fluorescence crossover around the threshold of the x-ray absorption.  Similar behavior has been observed before in FeTe \cite{hancock_evidence_2010}, though the origin of this crossover was not explained in detail.  We show that it stems from the finite lifetime of the core-hole present in the RIXS intermediate state and it represents a general feature in the RIXS spectra of metals.  We further provide a detailed analysis of the loss profiles, separating band structure features from many-body contributions.  In particular, for excitation energies well above the absorption edge, we observe a weak excitonic feature at about 1 eV energy loss that was not seen in previous studies.  We interpret this as an indirect contribution arising from electronic correlations within the Fe {\em d} bands.

Calculating the RIXS response of correlated metals is challenging because it requires the capability to handle multiple orbitals and large bandwidths, and to treat electronic correlations.  One electron methods, which are amenable to large bandwidths, are not well suited to accurately capture low energy excitations of correlated systems and have largely been restricted to core-to-core RIXS processes \cite{kas_rixs, Glover_Ge-RIXS}.  Here, we present a detailed, first-principles computational framework for evaluating and analyzing the valence electronic contributions to the RIXS spectra of metallic systems, taking the moderately correlated \BFA~as our test case.  Our initial approach solves the Bethe-Salpeter equation (BSE) for two-particle excitonic states based on an electronic band structure obtained from density functional theory (DFT).  Contrary to model Hamiltonians, cluster calculations, or approaches that reduce the electronic structure to Wannier orbitals of the correlated bands, the DFT-BSE includes all bands associated with all sites and orbital shells on an equal footing.  This reduces the arbitrariness of the method, avoids parameterization, naturally accounts for hybridization effects and is equally applicable to the transition metal and ligand edges.  The BSE has previously been applied to calculate the x-ray absorption and emission spectra of metals \cite{vinson_metals}, and the RIXS loss profiles of non-metals \cite{vinson_rixs, vorwerk_rixs, geondzhian_sto}, but never to the RIXS spectrum of a metal.

As we show in Sec.~\ref{sect:xasxes}, the Bethe-Salpeter equation, employing the typical static screening response, provides a reasonable accounting for the spectral features observed in the experimental x-ray absorption and emission spectra of \BFA.  However, certain discrepancies in the spectral intensity are evident.  This occurs because the BSE describes the excited state as a superposition of single electron-hole pairs and, due to the static screening approximation, fails to accurately capture spectral structures stemming from the generation of secondary electron-hole pairs.  We correct this deficiency in Sec.~\ref{sect:mnd} by generating, from first-principles, an effective spectral function for the excited state.  Convolving this spectral function with the BSE results yields spectra in excellent agreement with experiment.

Typically, calculations of RIXS spectra using one-electron approaches or the BSE formalism can account neither for secondary excitations in the direct RIXS channel nor for any part of the indirect RIXS contribution.  Starting from a simplified expression of the RIXS cross section, we show in Secs.~\ref{sect:fullrixs} and \ref{sect:model} that the effective excitonic spectral function enables us to calculate both of these responses.  This allows, for the first time, for a practical, first-principles method of calculating the RIXS response of metals and other periodic systems with full momentum dependence, including secondary electronic excitations and the indirect loss contribution.

\section{X-ray Absorption $\&$ Emission}
\label{sect:xasxes}

\begin{figure}
\includegraphics[trim=240 0 220 0,clip,width=8cm]{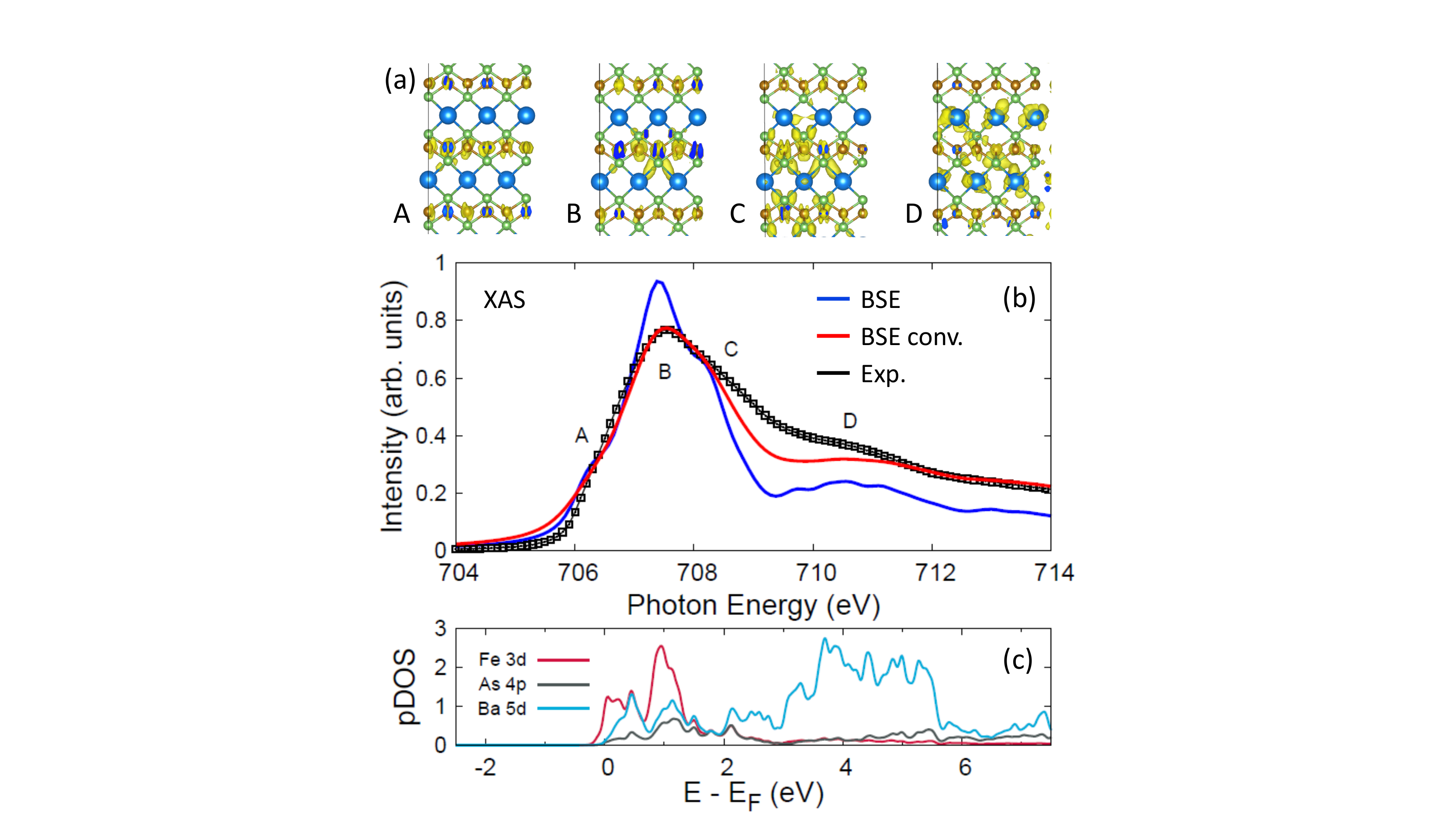}
\caption{\label{xas1} X-ray absorption of BaFe$_2$As$_2$ at the Fe L$_{3}$ edge.  (a) Photoelectron charge density isosurfaces for the excitonic states A-D indicated in the XAS spectrum.  (b) BSE calculation (blue curve) of the XAS compared with measurement (black symbols).  The red curve is obtained as described in the text by convolving the BSE spectrum with a Doniach-Sunjic spectral function of asymmetry parameter 0.14 that accounts for secondary electronic excitations. (c) Final state local projected density of states (pDOS) of the unoccupied levels for selected orbitals.}
\end{figure}

Before turning to the more complex RIXS results, we first consider the x-ray absorption (XAS) and x-ray emission (XES) spectra to better understand the RIXS intermediate and final states and their relations to the unoccupied and occupied electronic structure of BaFe$_2$As$_2$.  Figure \ref{xas1}b contains the measured and calculated Fe L$_{3}$ XAS.  The experimental spectrum displays a broad main peak with a maximum at an absorption energy around 707.7~eV accompanied by a shoulder region between about 710-711~eV. This lineshape is typical of metallic compounds and is rather similar to measurements performed on other Fe pnictides \cite{zhou_persistent_2013,pelliciari_intralayer_2016,kurmaev_identifying_2009,nomura_resonant_2016,yang_evidence_2009,pelliciari_presence_2016,pelliciari_local_2017,rahn_paramagnon_2019}. The good quality of the sample is evidenced by the lack of spurious distinct peaks associated with $\mathrm{Fe^{3+}}$ between 709-710~eV, which would be due to sample deterioration and oxidation.  

We calculate the XAS from first-principles by evaluating the two-particle excitonic Bethe-Salpeter equations within the OCEAN code \cite{gilmore_efficient_2015}.  The Bethe-Salpeter Hamiltonian consists of independent particle terms for the photoelectron and core-hole as well as an interaction kernel that includes the statically screened direct and bare exchange terms between the two particles \cite{shirley_bse2005}.  The independent particle contributions are obtained from a density functional theory (DFT) calculation of the ground-state electronic band structure.  The BSE interaction kernel is incorporated by summing ladder diagrams to infinite order.  The BSE calculation captures each of the features observed in the experiment, though with some discrepancies in intensities, which we address below.

The BSE calculations reveal at least three features within the main peak, labeled $A$, $B$, and $C$, and several features in the shoulder region, collectively labeled $D$.  To understand these contributions in more detail, we first consider in Fig.~\ref{xas1}c the final-state local projected density of states (pDOS) above the Fermi level around the absorbing Fe atom.  The pDOS suggests that the edge onset $A$ is mainly composed of Fe $3d$ and Ba $5d$ states, the main peak $B$ is dominated by Fe $3d$, hybridized to some extent with Ba $5d$ and As $4p$ states, the post-edge $C$ contains appreciable Ba $5d$ character and the shoulder $D$ is mainly derived from Ba $5d$ states.  While the final-state pDOS, evaluated in the presence of a core-hole on the absorbing Fe site, provides a valuable approximation of the orbital character of the final states, it does not account for weighting due to the transition matrix elements between the Fe 2$p$ core-level and the final states, nor for mixing of the final state excitons caused by multipolar terms of the Coulomb interaction.

To provide a more rigorous picture, we visually inspect the excitonic states associated with the main edge and shoulder regions.  For each incident photon energy ($A-D$), we obtain the two-particle, electron-hole wavefunction for the excitonic XAS final state.  Integrating out the core-hole coordinates leaves the wavefunction for the photoelectron.  Figure \ref{xas1}a presents isosurfaces for the charge density distributions corresponding to these photoelectron wavefunctions.  These orbital isosurfaces mainly confirm the pDOS interpretation, except that they indicate the pre-edge feature $A$ has very little Ba character, contrary to the pDOS prediction, and the shoulder region maintains more Fe character than implied in the pDOS.  Consistent with the metallic nature of BaFe$_2$As$_2$, the photoelectrons are not strongly localized around the absorption site but are rather diffuse and delocalized.  At the main peak, the photoelectron is largely confined within the first iron-arsenide layer, consisting of hybridized Fe $3d$ and As $4p$ states.  In the post-edge and shoulder regions, the photoelectron has a greater spatial expanse and gains appreciable $5d$ orbital character on the Ba sites.

\begin{figure}
\includegraphics[trim=240 0 220 0,clip,width=8cm]{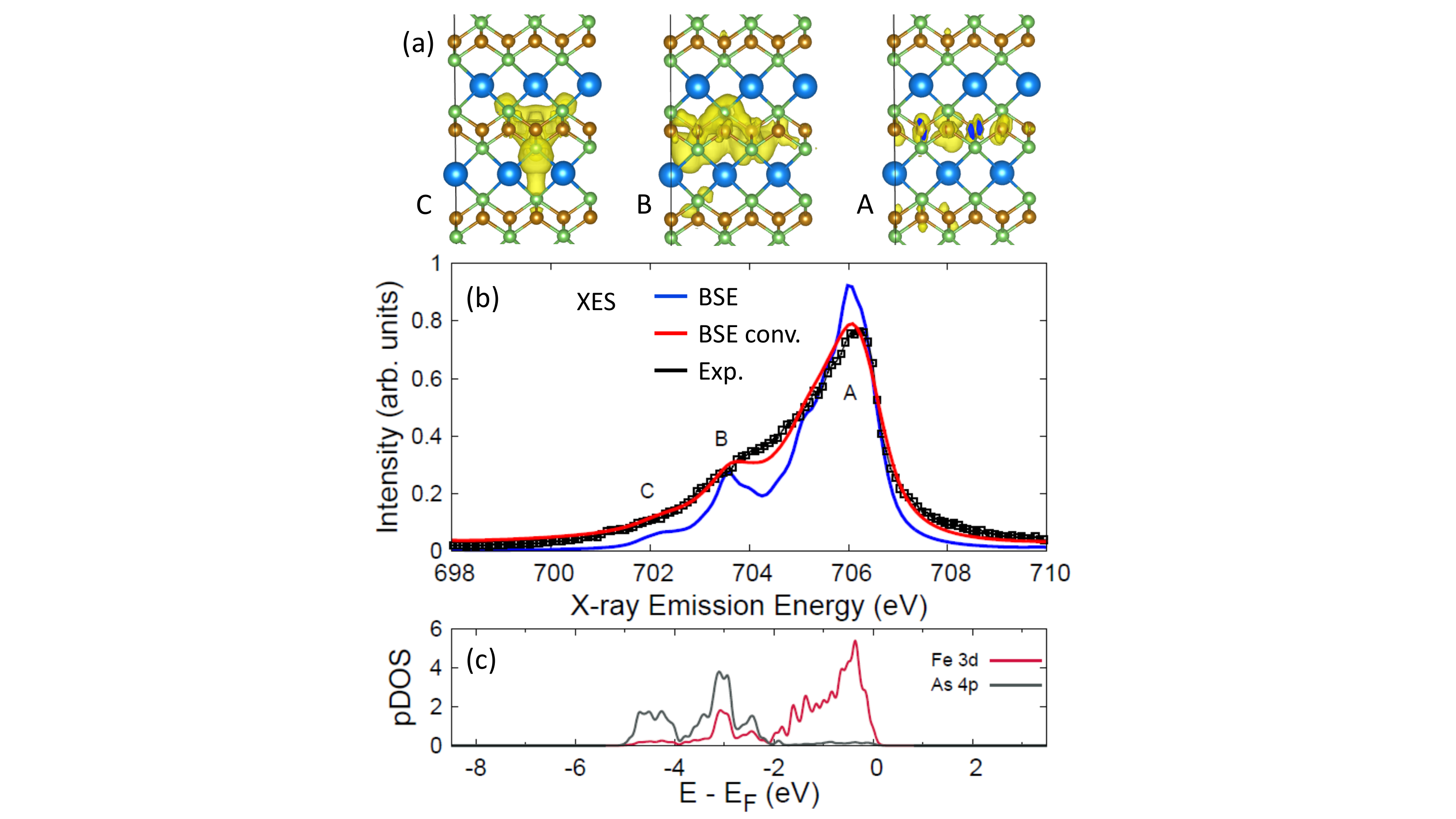}
\caption{\label{xes1} X-ray emission of BaFe$_2$As$_2$ at the Fe L$_{3}$ edge.  (a) Orbital isosurface plots of the valence-hole contribution for the final-state of the emission process corresponding to emission energies A-C indicated in the main figure.  (b) BSE calculated XES (blue curve) compared with measurement (black symbols).  The red curve is obtained as described in the text by convolving the BSE spectrum with a Doniach-Sunjic spectral function of asymmetry parameter 0.07 that accounts for secondary electronic excitations. (c) Ground-state projected density of states for the occupied states of selected orbitals.}
\end{figure}

Figure \ref{xes1}b presents the measured and calculated XES at the Fe L$_3$ edge.  The experimental spectrum consists of a broad, asymmetric emission line, however, the calculation reveals finer structure with a main feature $A$ at around 706 eV, a distinct second feature $B$ near 703.5 eV, and a weak peak $C$ at 702 eV.  Figure \ref{xes1}c gives the ground-state pDOS of the occupied states just below the Fermi level, showing that the main feature $A$ consists entirely of Fe $3d$ orbitals, the shoulder feature $B$ is comprised of Fe $3d$ $-$ As $4p$ hybridized states, and the weak feature $C$ at 702 eV is mainly of As $4p$ character.  This is further supported by the orbital plots of the part of the excitonic wavefunctions corresponding to the final state valence hole, which are shown in Fig.~\ref{xes1}a. To further corroborate these assignments, we note that peaks $B$ and $C$, which derive appreciably from As orbitals, are absent in emission measurements of FeTe \cite{hancock_evidence_2010}.

\begin{figure}
\includegraphics[width=8cm]{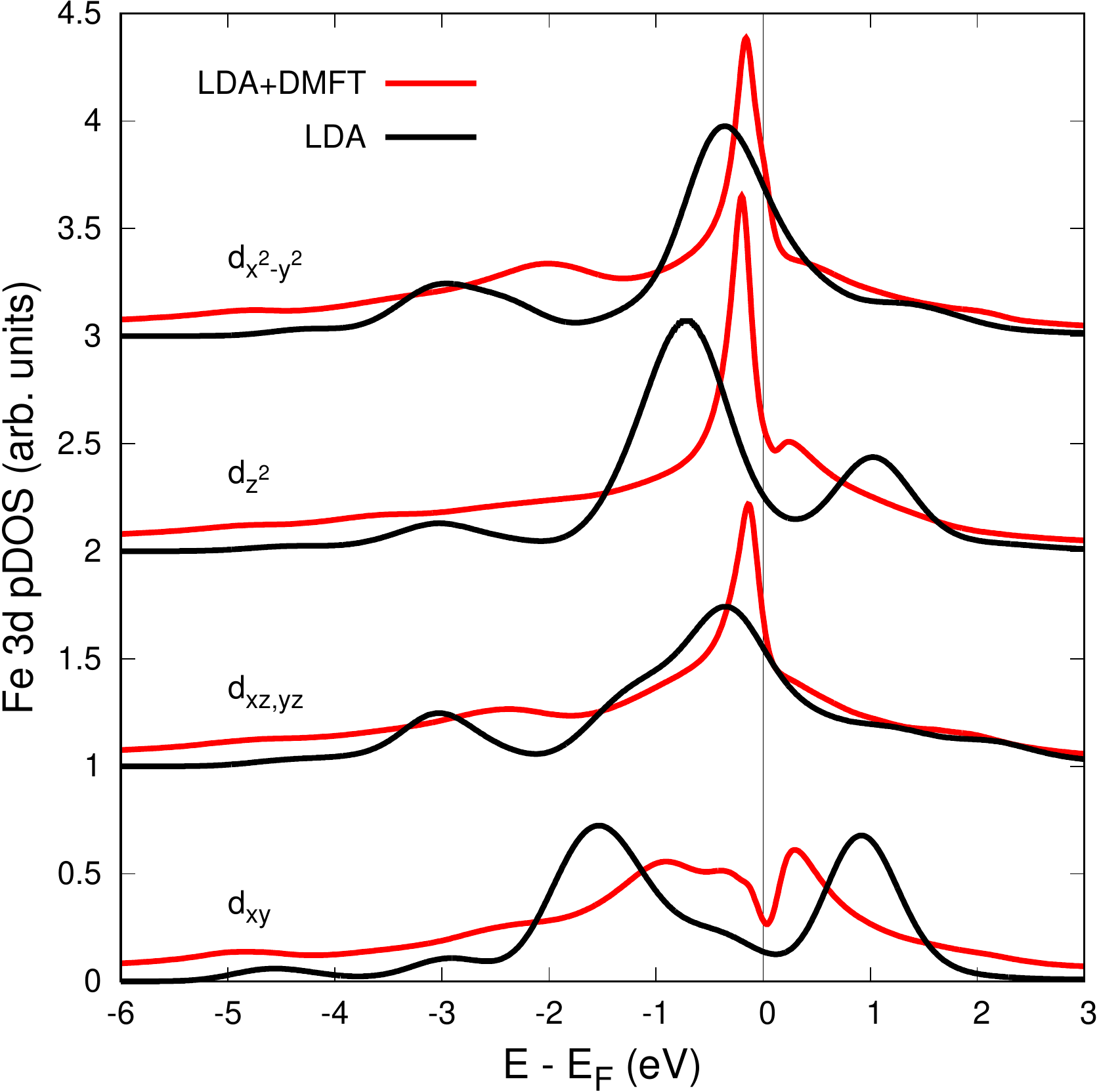}
\caption{\label{pDOS} Fe $3d$ pDOS for \BFA~calculated within LDA density functional theory (black) and dynamical mean field theory based on LDA-DFT (red).  Curves are offset vertically for clarity.}
\end{figure}

The BSE spectra are in reasonable agreement with the experimental results, though they are somewhat overstructured. The calculated XAS is too sharp at the main peak and lacks intensity in the post-edge and shoulder regions. The comparison of the calculated to measured XES is similar.  BaFe$_2$As$_2$ is considered a moderately correlated electron system \cite{Skornyakov_PhysRevB.80.092501}, raising doubts \cite{Derondeau_SciRep_2017} about the appropriateness of basing spectral calculations on a DFT electronic structure obtained with the local density approximation (LDA) to the exchange-correlation functional.   However, the energies of all features correspond well between the calculated and measured XAS, suggesting that the underlying LDA band structure serves as a reasonable starting point for the BSE calculations.  To further alleviate this concern, in Fig.~\ref{pDOS} we compare the Fe $3d$ pDOS between an LDA-DFT calculation and a dynamical mean field theory (DMFT) calculation.  The LDA-DMFT calculation tends to sharpen the spectral function about the Fermi level, but generally shifts spectral peaks by less than 1 eV.  These changes on their own do not appear to explain the reduced intensity with respect to the experiment in the shoulder region of the XAS, which is primarily associated with the Ba $5d$ orbitals.

The deviations in intensities of the spectral features could still be due to an inaccurate description of the spatial extent of the LDA Kohn-Sham wavefunctions, however, it is likely that a larger effect originates in the fundamental two-particle description of the BSE that neglects secondary electronic excitations beyond what is captured by the static screening response.  The BSE Hamiltonian $H_{\alpha c, \alpha^{\prime} c^{\prime}}^{BSE}$ is expressed in a two-particle, electron-hole basis.  For core-level absorption, $\alpha$ denotes both the $m_l$ and $m_s$ values for the core-hole while $c$ incorporates the band index $n$, wavevector ${\bf k}$ and spin $\sigma$ of the conduction-level photoelectron.  The excitonic eigenstates $\Omega^{\lambda}$ are obtained by solving $\sum_{\alpha^{\prime} c^{\prime}} H_{\alpha c, \alpha^{\prime} c^{\prime}}^{BSE} \Omega_{\alpha^{\prime} c^{\prime}}^{\lambda} = E^{\lambda} \Omega_{\alpha c}^{\lambda}$.  These eigenstates are single Slater determinants that can be expressed as linear superpositions of electron-hole pairs such that $\Omega^{\lambda} = \sum_{\alpha c} C_{\alpha c}^{\lambda} \ket{\alpha,c}$.  The true many-body excited-state will be more complex and involve the generation of additional, valence-conduction electron-hole pairs as the charge density responds to the creation of the core-hole.  We refer to these as the secondary excitations neglected within the BSE.  Physically, these secondary excitations correspond to multiplet features in open-shell systems, as well as shake-up or shake-off satellites and even plasmon satellites.

It is possible to go beyond the single electron-hole pair description of the Bethe-Salpeter equation and effectively account for secondary excitations by expressing the many-body absorption coefficient 
~
\begin{equation}
\label{convolution}
    \mu(\omega) = \sum_{\lambda} \int d\omega^{\prime} A_{\lambda}(\omega^{\prime}) \mu_{\lambda}^{0}(\omega - \omega^{\prime}) \,
\end{equation}

\noindent as a convolution of the single exciton BSE XAS spectrum $\mu_{\lambda}^{0}$ with a spectral function $A_{\lambda}$ that accounts for these secondary excitations \cite{kas_eh-cumulant_2016, woicik_k-edges, woicik_l-edges}.  The spectral function $A_{\lambda}$ is specific to a particular excitonic eigenstate $\Omega^{\lambda}$ of the BSE Hamiltonian.  X-ray photoemission studies of \BFA~observed that the structure of the Fe 2$p$ core-hole spectral function is asymmetric and can be reproduced very well with a Doniach-Sunjic (DS) lineshape \cite{de_jong_high-resolution_2009}.  This motivates the choice of a Doniach-Sunjic profile to approximate the effective spectral function $A_{\lambda}(\omega)$ in Eq.~\ref{convolution}.  In Sec.~\ref{sect:mnd}, we will validate this assumption by explicitly constructing the spectral function from first-principles.  The expression for the DS lineshape, and the meaning of the parameters, are provided in Appx.~\ref{append:comp.details}. 

The XAS profile resulting from Eq.~\ref{convolution} (see the red curve in Fig.~\ref{xas1}b) agrees much better with the experimental result than the bare BSE spectrum.  Not only does the main peak trace the experimental profile very well, spectral weight is shifted from the main peak to the post-edge region leading to a significant improvement in the post-edge and shoulder features.  A small amount of spectral weight is still missing in the immediate post-edge area around 709 eV.  These small differences do not arise from experimental broadening of the XAS spectrum as the energy resolution of the measurement is less than the core-hole lifetime.  The differences could potentially indicate minor shortcomings from the underlying LDA calculation.  Alternatively, it is possible that the profile of secondary excitations is more complex than suggested by the Doniach-Sunjic lineshape.  In this case, local correlation effects on the Fe site could produce a more prominent $dd$ excitation in the range of 1-2 eV.  We consider this possibility in more detail in Sec.~\ref{sect:model}.

Similarly to the XAS results, the BSE calculation for the emission spectrum is sharper than the measured spectrum with missing spectral weight below the main edge.  Convolution of the BSE spectrum with a DS lineshape again leads to considerably improved agreement with the experimental result, indicating that the emission process, {\em i.e.} filling the core-hole, also incoherently kicks up various low energy electron-hole pairs.  However, the Doniach-Sunjic asymmetry parameter for emission (0.07) is smaller than for absorption (0.14), suggesting that the effect, while still significant, is less pronounced in the emission case.  This is potentially explicable in that for XES, both initial and final states have a hole and no photoelectron, reducing the change in the potential between initial and final states compared to the XAS process.

\section{Resonant inelastic X-ray scattering}
\label{sect:rixs}

The second order approximation to the RIXS loss profile can be expressed in Green's function notation as
~
\begin{equation}
\label{rixsloss}
    \sigma(\omega_i, \omega_o) \propto -{\rm Im} \bracket{0}{\hat{d}_{i}^{\dagger}G(\omega_i)\hat{d}_{_o}G(\omega_{l})\hat{d}_{_o}^{\dagger}G(\omega_i)\hat{d}_{_i}}{0} \, .
    %\sigma(\omega_i, \omega_o) = -\frac{1}{\pi}{\rm Im} \bracket{0}{\hat{d}_{\hat{\epsilon}_i}^{\dagger}G^{(2)}(\omega_i)\hat{d}_{\hat{\epsilon}_o}G^{(2)}(\omega_{loss})\hat{d}_{\hat{\epsilon}_o}^{\dagger}G^{(2)}(\omega_i)\hat{d}_{\hat{\epsilon}_i}}{0} \, .
    %\sigma(\omega_i, \omega_o) = -\frac{1}{\pi}{\rm Im} \bracket{0}{\hat{d}_{\hat{\epsilon}_i}^{\dagger}G(\omega_i)\hat{d}_{\hat{\epsilon}_o}G(\omega_{l})\hat{d}_{\hat{\epsilon}_o}^{\dagger}G(\omega_i)\hat{d}_{\hat{\epsilon}_i}}{0} \, .
\end{equation}

\noindent The photon operator $\hat{d}_i$ ($\hat{d}_o$) is associated with the incident (outgoing) photon, which has polarization vector $\hat{\epsilon}_i$ ($\hat{\epsilon}_o$), wavevector ${\bf k}_i$ (${\bf k}_o$), and energy $\omega_i$ ($\omega_o$).  In the present work we take the dipole approximation $\hat{d}=\hat{\epsilon}\cdot {\bf r}$.  The energy and momentum transferred during the scattering process are $\omega_{l} = \omega_i - \omega_o$ and ${\bf q} = {\bf k}_i - {\bf k}_o$.  The first and last Green's functions propagate the intermediate state in the presence of the core-hole while the central Green's function propagates the final, valence-excited state.

%\begin{widetext}
Within the Bethe-Salpeter framework, the evaluation of Eq.~\ref{rixsloss} for the RIXS loss profile is separated into three steps.  This begins by generating the XAS-like intermediate core-excited state
~
\begin{equation}
    \ket{y(\omega_i,\hat{\epsilon}_i)} = G(\omega_i) \hat{d}_i \ket{0} \, .
\end{equation}

\noindent The BSE reduces the many-body problem to an effective two-particle, electron-hole basis consisting of Fe $2p$ core levels $\alpha$ and the unoccupied conduction states $c$.  Approximating the propagator as $G(\omega) = [\omega-H^{BSE}]^{-1}$, within this two-particle basis the RIXS intermediate (core-excited) state is 
\begin{widetext}
\begin{equation}
 \label{yvec}
    %\ket{y(\omega_i,\hat{\epsilon}_i)} = \sum_{\alpha c} \sum_{\alpha^{\prime} c^{\prime}} \ket{\alpha^{\prime} c^{\prime}} \bracket{\alpha^{\prime} c^{\prime}}{\frac{1}{H_{BSE}-\omega_i}}{\alpha c} \bracket{\alpha c}{\hat{d}_i}{0} \, .
    \ket{y(\omega_i,\hat{\epsilon}_i)} = \sum_{\alpha c, \alpha^{\prime} c^{\prime}} \ket{\alpha^{\prime} c^{\prime}} \bracket{\alpha^{\prime} c^{\prime}}{\frac{1}{\omega_i - H^{BSE}}}{\alpha c} \bracket{\alpha c}{\hat{d}_i}{0} \, .
\end{equation}

\noindent Evaluation of Eq.~\ref{yvec} requires solving a core-level BSE problem.

The second step of the RIXS calculation constructs the x-ray emission state
~
\begin{equation}
    \ket{x(\omega_i,\hat{\epsilon}_i,\hat{\epsilon}_o)} = \hat{d}_o^{\dagger} \ket{y(\omega_i,\hat{\epsilon}_i)} \, .
\end{equation}

\noindent This is expressed within the two-particle description as
~
\begin{equation}
    \ket{x(\omega_i,\hat{\epsilon}_i,\hat{\epsilon}_o)} = \sum_{v} \sum_{\alpha c} \ket{v c} \bracket{v c}{\hat{d}_o^{\dagger}}{\alpha c} \braket{\alpha c}{y(\omega_i,\hat{\epsilon}_i)} 
\end{equation}

\noindent where $v$ runs over previously occupied valence states and represents a band index, wavevector and spin direction.  The vector $\ket{x(\omega_i,\hat{\epsilon}_i,\hat{\epsilon}_o)}$ contains a valence-conduction electron-hole pair, though this state is generally not an eigenstate of the RIXS final state.  The final-state eigenstates and their corresponding RIXS intensities are obtained in the last step of the RIXS calculation by evaluating
~
\begin{equation}
    \sigma_{\bf q}(\omega_i,\omega_o) \propto -{\rm Im} \bracket{x(\omega_i,\hat{\epsilon}_i,\hat{\epsilon}_o)}{\frac{1}{\omega_{l} - H_{\bf q}^{BSE}}}{x(\omega_i,\hat{\epsilon}_i,\hat{\epsilon}_o)}
\end{equation}

\noindent where this is now a valence-level BSE problem evaluated with finite momentum transfer ${\bf q}$.  We obtain the RIXS spectra using the OCEAN code \cite{shirley_rixs,gilmore_efficient_2015,vinson_rixs}.

\end{widetext}

It is typical to classify the RIXS response in terms of direct and indirect contributions.  However, there is not a clear consensus as to the definition of these terms.  From an empirical perspective, one is inclined to classify the fluorescence contribution as direct RIXS and Raman features, having constant energy loss, as the indirect contribution.  This is qualitatively consistent with the canonical figures presented by Ament {\em et al.}~\cite{ament_resonant_2011}.  In that work, the authors construct a formal definition of the direct and indirect terms by expressing the intermediate-state propagator as the Dyson equation $G = G_{0} + G_{0} H_{C} G$ where $G_{0}$ is the ground-state propagator and $H_{C}$ introduces the core-hole (core-exciton) perturbation.  They associate the direct term with the ground-state propagator $G_{0}$ and the indirect term with $G_{0} H_{C} G$.  Within this definition the direct contribution should be calculated neglecting all excitonic effects while our BSE approach would account for both direct and indirect responses since we perform calculations using the full propagator, albeit an approximation thereof.

The above definitions were presented largely with Mott insulators in mind.  At least when considering metals or systems for which the gap is small compared to the correlated orbital bandwidth, we prefer to introduce an alternative, physically motivated definition of the direct and indirect RIXS contributions.  The objective is to separate band structure contributions, giving the direct signal, from the correlated response of the system to a perturbation, captured by the indirect contribution.  Beginning with an independent electron approximation for the ground-state, we define the direct contribution as that associated with excitation of a core-electron to a state above the Fermi level and decay of another electron from a state below the Fermi level, filling the core-hole.  The populations of all other levels remain unaffected, though excitonic effects such as screening of the core-hole imply that the energies of all levels shift, modulating oscillator intensities and differentiating our definition from that offered by Ament {\em et al.}~\cite{ament_resonant_2011}.  The indirect contribution arises when screening of the core-level exciton, in addition to shifting the energies of all the states, also effects their populations through the generation of secondary excitations, {\em e.g.} shake-up or shake-off processes.  Collective excitations, such as plasmons, magnons and phonons could also appear in the indirect channel.  The pure indirect channel derives from the recombination of the original photoelectron with the core-hole.  However, direct and indirect effects are not cleanly separable and direct-channel processes may be accompanied by indirect-channel losses.
%[~\citenum{ament_resonant_2011}]

Our redefinition associates the direct contribution with the primary core-hole--photoelectron exciton, treated at the quasiparticle level and neglecting any secondary excitations.  This is precisely what is evaluated by the BSE calculations taking the usual static screening approximation.  Indirect effects arise from all secondary valence-conduction excitations caused by the perturbation of the primary core-exciton.  Within the context of XAS and XES, we accounted for these secondary excitations with the excitonic spectral function in Eq.~\ref{convolution}.  The BSE calculations fail to capture any of the indirect RIXS contribution since the intermediate state is explicitly restricted to a single core-hole plus single conduction electron basis.  However, in Sec.~\ref{sect:fullrixs}, we use the excitonic spectral function to develop an effective extension for implicitly including the indirect channel.

\begin{figure*}
\centering
\includegraphics[width=10cm,angle=270]{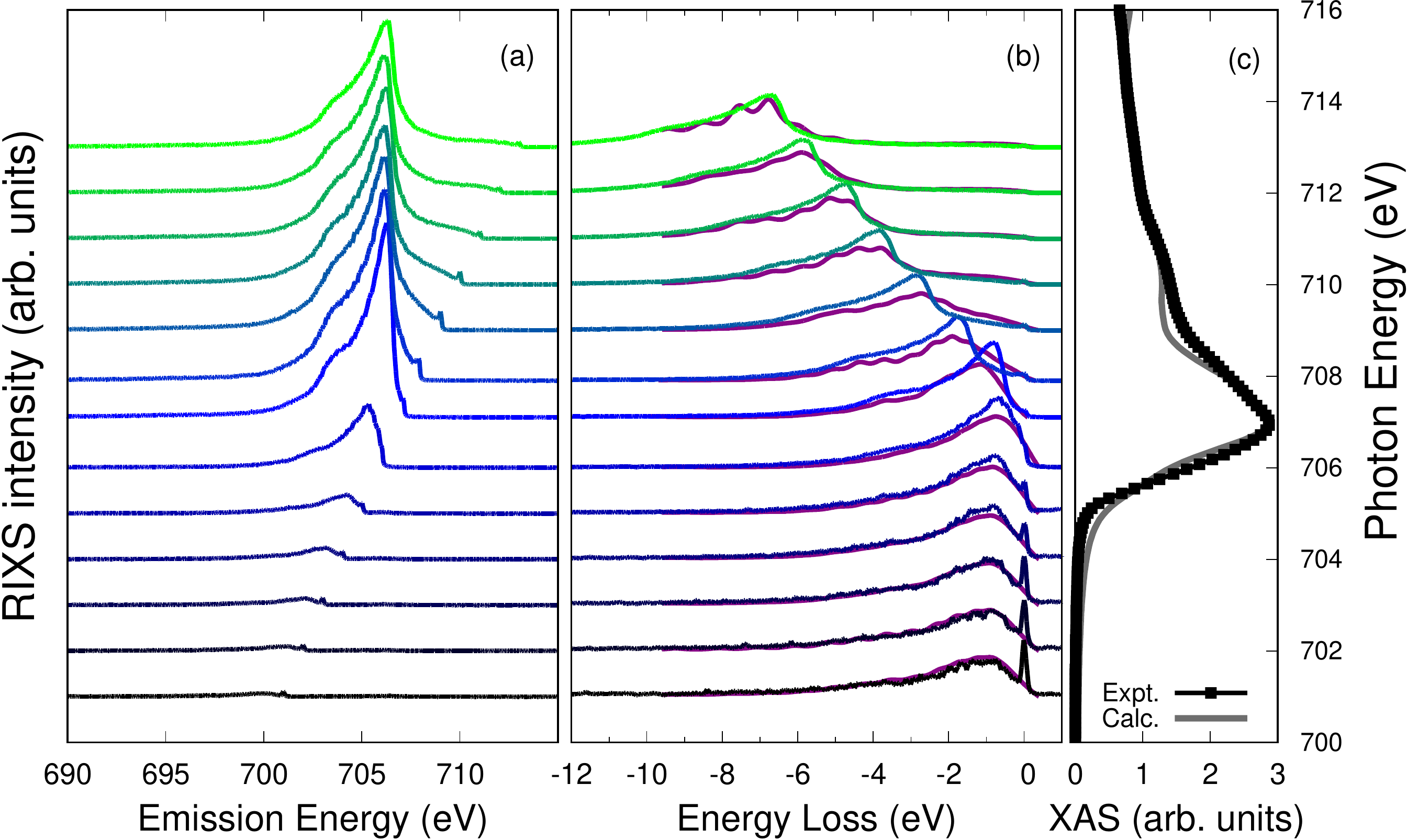}
\caption{\label{rixs1} (a) Experimental RIXS spectra recorded at the Fe $L_3$ edge of \BFA~plotted versus emitted photon energy.  (b) Normalized experimental and calculated (purple curves) RIXS spectra plotted versus energy loss.  In both (a) and (b) curves are staggered vertical to align with their corresponding incident photon energy with respect to the Fe $L_3$ XAS profile (c).  The momentum transfer for all spectra, experimental and calculated, was (0.31, 0,31, 0) in reciprocal lattice units.}
\end{figure*}

Figure \ref{rixs1}b compares the experimental and calculated RIXS spectra for several incident photon energies, plotted as a function of energy loss.  Each loss curve is individually area normalized in order to better compare features in the loss profiles across the different incident photon energies.  Compared to previous RIXS studies on non-metallic or less itinerant transition metal oxides \cite{fabbris_orbital_2016,minola_collective_2015,dean_itinerant_2014,bisogni_ground_2016,meyers_doping_2017,dantz_quenched_2016,ellis_correlation_2015,perret_coupled_2018,fatale_electronic_2017,mcnally_electronic_2019,Schmitt_V2O3}, the loss peak for \BFA~is not sharp but rather broad in energy.  The width in energy of the peak is comparable to the occupied electronic bandwidth, suggesting that the nature of this peak is different from conventional, indirect \textit{dd} excitations, and more a reflection of the direct band structure contribution. 

The most evident trend is the Raman-to-fluorescence crossover as the incident photon energy traverses the XAS peak.  For incident energies tuned below the XAS threshold, the loss profile peaks around 1 eV and extends out several eV\textit{s} to higher energy loss.  This feature behaves in a Raman-like way, {\it i.e.} its position in energy loss is independent of the incident photon energy.  However, as the incident photon energy passes the threshold of the x-ray absorption spectrum, the RIXS loss profile begins to behave in a fluorescence-like fashion, {\it i.e.} the features shift outward in energy loss proportionally with the increase of the incident photon energy.  This fluorescence behavior is more clearly illustrated in Fig.~\ref{rixs1}a in which the RIXS loss profiles are plotted versus absolute emitted photon energy (the spectra are not area renormalized in Fig.~\ref{rixs1}a).  In this presentation, the RIXS spectra appear at constant emitted photon energy above the XAS threshold.  This kind of Raman-to-fluorescence crossover was previously observed for FeTe \cite{hancock_evidence_2010}.

%\textcolor{green}{
%The electrons leading to the Raman behaviour are the same valence electrons leading to the fluorescence, implying that the observation of a Raman-like mode can not describe the itinerant or localized nature of a system \cite{monney_mapping_2012}.}

The calculations reproduce well the experimental trends, capturing both the specific loss profiles and the overall Raman-to-fluorescence crossover as the incident photon energy is scanned across the absorption threshold.  Fluorescence behavior is often associated with direct RIXS processes and Raman behavior with indirect processes.  In this view, it is surprising that the calculations, which only include direct RIXS processes, capture both fluorescence and Raman behavior.  We demonstrate now that the overwhelming contribution to the experimental loss spectra in both the fluorescence and Raman regions correspond to direct RIXS processes and that the Raman-to-fluorescence behavior can be explained within the direct RIXS framework as a core-hole lifetime effect.

The main features of the loss profile appear consistent between the different incident energies and bear a strong resemblance to the x-ray emission spectrum from Fig.~\ref{xes1}.  For all incident photon energies, there is a main loss peak followed after another 2-3 eV of loss by a shoulder feature.  A much weaker third feature can be seen about 1-2 eV further in energy loss.  There are minor differences in relative peak intensities as the incident photon energy is varied, but all loss profiles clearly reflect the XES.  This indicates that the RIXS loss profile is dominated by direct emission from occupied valence levels as opposed to the generation of secondary $dd$ or charge-transfer excitations.  This description holds not just for the fluorescence-like region, but also in the Raman-like region.

\begin{figure*}
\includegraphics[width=17cm]{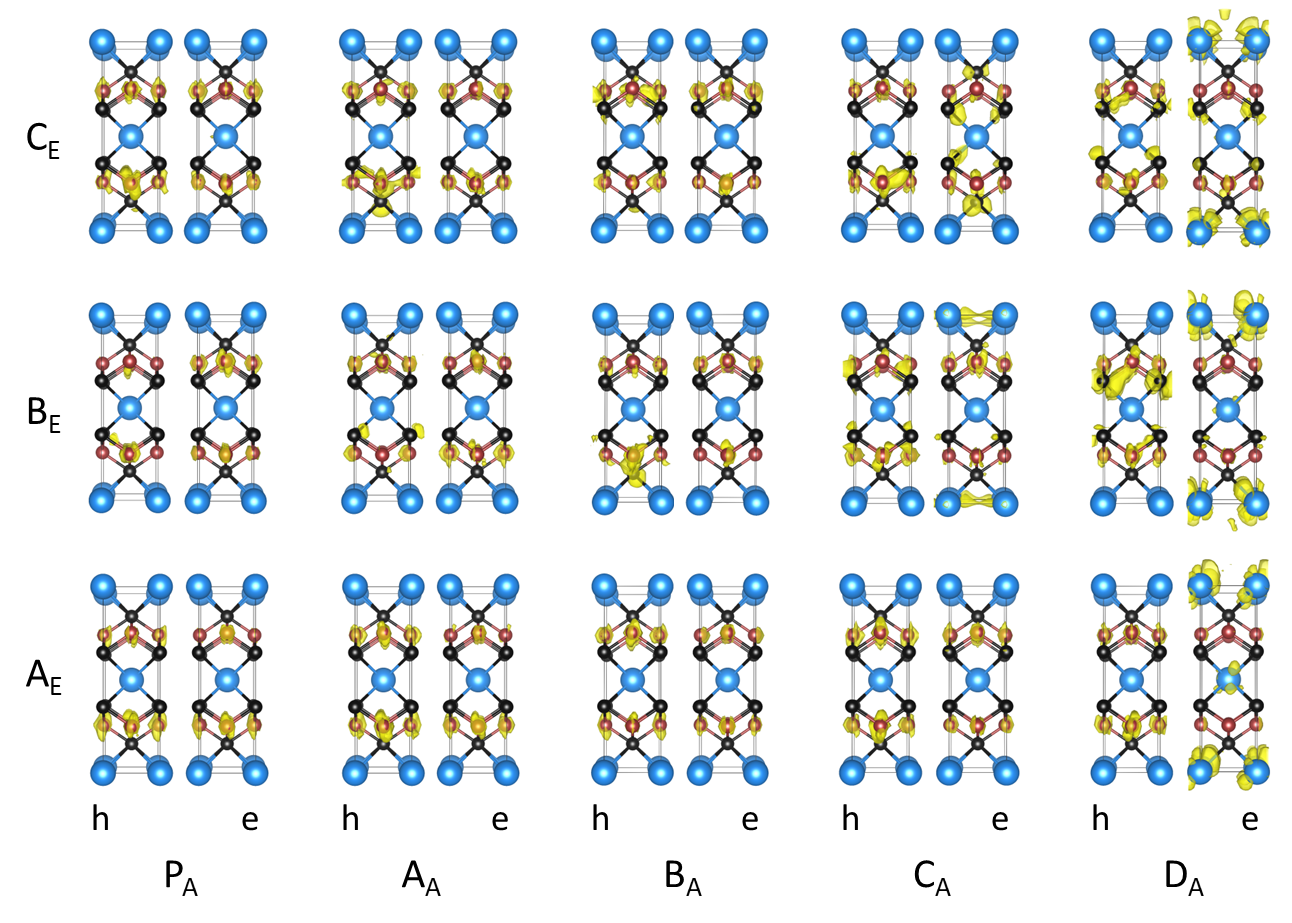}
\caption{\label{rixs2} Isosurface plots for the charge density of the hole (h) and electron (e) components of the RIXS final state exciton for different combinations of absorption feature ($P_{A},A_{A},B_{A},C_{A},D_{A}$) and emission feature ($A_{E},B_{E},C_{E}$); see labels in Figs.~\ref{xas1}b and \ref{xes1}b, respectively.  Iron atoms are shown in red, arsenic in black, and barium in blue.}
%At and below the absorption threshold, the charge density distribution of the excited electron is largely independent of both the incident and emitted photon energies, whereas the spatial extent of the electron charge distribution expands for incident photon energies above threshold ($C_{A},D_{A}$).  For all incident photon energies, the charge density distribution of the excited electron is largely independent of the emission energy, indicating that excitonic effects are minimal in the final state.  Likewise, the contour plots for the final-state valence-level hole are largely independent of the incident photon energy.}
\end{figure*}

To aid in understanding the Raman-to-fluorescence crossover in more detail, Fig.~\ref{rixs2} presents the electron and hole charge density contributions to the RIXS final-state, valence-conduction excitons for several combinations of incident and outgoing photon energies.  In particular, we consider incident photon energies corresponding to the features $\{A,B,C,D\}$ of the x-ray absorption spectrum, which we label as $\{A_A,B_A,C_A,D_A\}$, and outgoing photons associated with emission from features $\{A,B,C\}$ of the x-ray emission spectrum, which we label as $\{A_E,B_E,C_E\}$.  Additionally, we also include one pre-edge incident photon energy, below the XAS threshold, labeled $P_{A}$, corresponding to an incident photon energy of 705 eV.  For incident photon energies below threshold, and to a slightly lesser extent above threshold ($C_A$ and $D_A$), the orbital plots of the hole state in the valence band are approximately independent of the incident photon energy for each of the three main features in the RIXS loss profiles.  For incident photons tuned from below threshold to the peak of the absorption spectrum ($P_A$, $A_A$ and $B_A$), the final-state electron charge density is largely confined within the FeAs layer.  As the incident photon energy increases above the peak and into the tail of the $L_3$ edge, the electron charge density shifts more to the Ba layers.  This mirrors closely the behavior of the excitonic states of the x-ray absorption spectrum in Fig.~\ref{xas1}.  Furthermore, this indicates that the RIXS loss spectra are dominated by incoherent emission processes for all incident energies and that, due to the metallic nature of the sample, excitonic effects are weak in the valence-excited RIXS final-state.

For incident photon energies below the XAS threshold, the electron component of the exciton orbitals in Fig.~\ref{rixs2} indicates that the excited electron in the RIXS final state occupies the same orbitals within the FeAs plane, just above the Fermi level.  At the same time, for a given feature in the RIXS loss profile, the valence hole orbitals are also independent of the incident photon energy.  Thus, at each incident photon energy below threshold, the RIXS final states consist of the same electron-hole pairs, generated through a direct scattering process.  While it may seem that an incident photon tuned below threshold lacks sufficient energy to excite a core electron above the Fermi level, the finite core-hole lifetime allows the excitation with a probability that diminishes as the photon energy is further detuned below threshold.  Consistent with this explanation of core-hole broadening allowed transitions, the intensities of the RIXS spectra rapidly attenuate below the absorption threshold when properly scaled, as in Fig.~\ref{rixs1}a.  In Appx.~\ref{append:lifetime}, we show that this attenuation becomes more pronounced as the core-level lifetime broadening decreases.  This direct RIXS Raman-to-fluorescence crossover should be a common feature among metallic samples and we anticipate that it will also be observed in non-metallic materials when the core-hole broadening is similar to or larger than the energy gap.

\section{Secondary excitations in metals}
\label{sect:mnd}

\begin{figure}
\includegraphics[width=8.5cm,trim={1cm 0 0 0},clip]{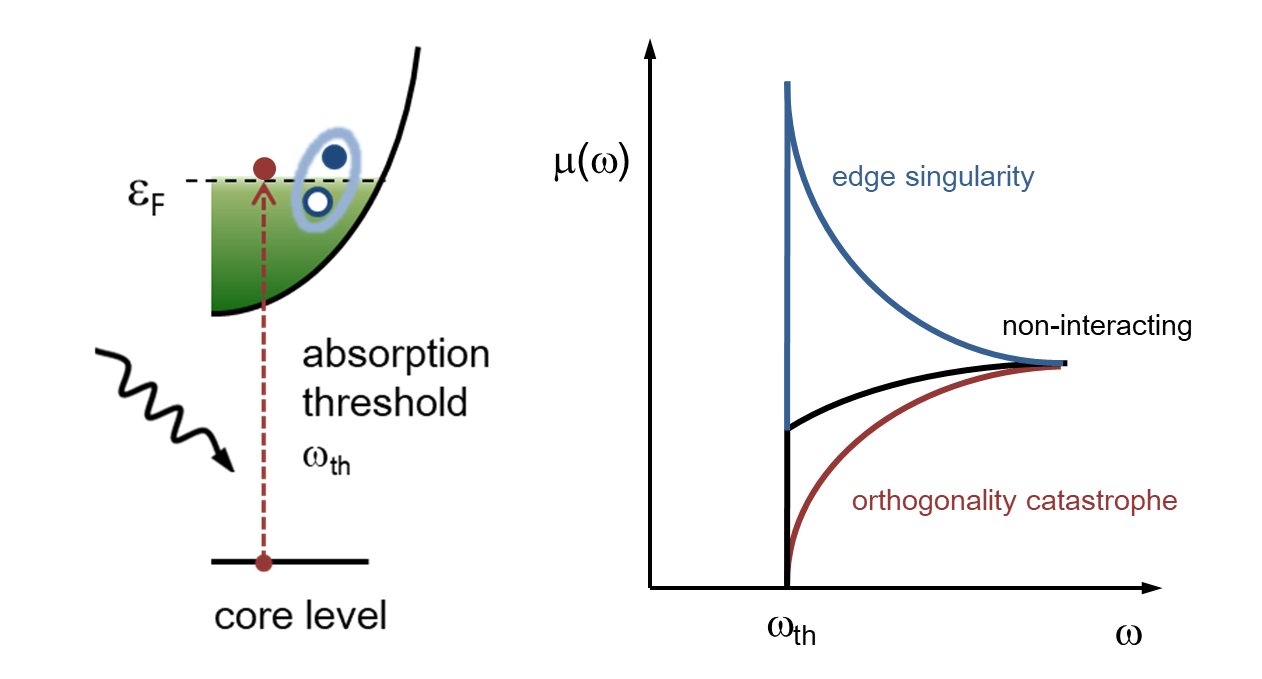}
\caption{\label{fig:mnd} Schematic of the core-level absorption edge profile in metals.  The Anderson orthogonality catastrophe acts to quench absorption at threshold, $\omega_{th}$, while the Mahan edge singularity leads to a divergence at threshold due to the generation of secondary excitations of vanishing energy.  Both profiles follow a power-law behavior.  These competing effects lead to a finite absorption at threshold with either the rounded or peaked nature of the edge being determined by the scattering phase shifts.}
\end{figure}

The theoretical study of the response of a metallic system to the sudden creation of a core-hole, due to the absorption of a photon, was initially investigated at the end of the '60\textit{s} in what has become known as the Mahan-Nozi\'eres-de Dominicis (MND) model \cite{anderson_infrared_1967, mahan_excitons_1967, nozieres_singularities_1969}. The MND model accounts for many-body effects and the generation of secondary excitations by synthesizing two competing effects: the Anderson orthogonality catastrophe \cite{anderson_infrared_1967} and Mahan's edge singularity \cite{mahan_excitons_1967}.  The orthogonality catastrophe states that the many-body final state (in the presence of the core-hole) is orthogonal to the initial (ground) state and therefore the transition intensity should go to zero at threshold.  Meanwhile, the edge singularity predicts an infinite transition probability due to the generation of secondary electron-hole pairs that can have vanishing energy at the Fermi level in a metal.  The MND theory combines the two ideas, yielding a finite absorption intensity at threshold.  Figure \ref{fig:mnd} depicts these processes schematically.

The Mahan response of a material can generate both collective plasmons and incoherent electron-hole pairs.  In metallic systems, electron-hole pairs often dominate due to their vanishing energy cost.  Based on MND theory, the x-ray photoemission (XPS) lineshape of metals was studied by Doniach and Sunjic who provided a description of their typical asymmetric XPS profile \cite{doniach_many-electron_1970}.  This consists of a divergent quasiparticle peak with a smooth tail that accounts for the secondary electron-hole pairs and falls off as a power-law.  In previous work \cite{de_jong_high-resolution_2009}, the Fe $2p$ XPS of \BFA~was fit with a DS lineshape yielding an asymmetry parameter of 0.44, which is identical to the value obtained for elemental iron \cite{fanelsa_fe-xps_1996}, suggesting that strong electron correlations do not play a significant role in the response to the core-hole.

In Sec.~\ref{sect:xasxes}, we demonstrated the efficacy of applying MND theory to go beyond the single electron-hole pair description of the Bethe-Salpeter equation by utilizing effective spectral functions to account for secondary excitations.  This was accomplished by expressing in Eq.~\ref{convolution} the total absorption or emission spectrum as a convolution of the BSE spectrum and a spectral function.  In this section, we explicitly construct the XAS spectral function from first-principles and confirm that the DS lineshape is a reasonable approximation.

Since we evaluate the bare absorption spectrum within the two-particle BSE framework, the spectral function $A$ quantifies the response of the rest of the system to the creation of an exciton.  Therefore, the spectral function corresponds to the imaginary part of the Green's function for the exciton.  To generate a spectral function with suitable structure, we express the Green's function in the cumulant representation $G(t)=G_0(t) e^{C(t)}$ and follow the work of Kas {\em et al.}~to construct the cumulant $C(t)$ from first-principles \cite{kas_rt-cumulant_2015, kas_eh-cumulant_2016, woicek_ct-sto_2020}.  The bare two-particle Green's function for the exciton $G_0$ is obtained from the solution of the BSE as discussed above.  To second order, the cumulant is 
~
\begin{equation}
    C(t) = \int d\omega \frac{\beta(\omega)}{\omega^2} \left ( e^{-i\omega t} + i\omega t - 1 \right )
\end{equation}

\noindent where the quasi-boson excitation spectrum for the secondary electronic excitations is given by
~
\begin{equation}
    \beta(\omega) = \sum_{q,q^{\prime}} V_{q}^* V_{q^{\prime}} {\rm Im} [\chi(q,q^{\prime};\omega)] \, .
\end{equation}

\noindent The charge density response function is 
~
\begin{equation}
    \chi(q,q^{\prime};\omega) = i\int dt \, e^{i\omega t} \langle \rho_{q}(t) \rho_{q^{\prime}}(0) \rangle \theta(t)
\end{equation}

\noindent and $V_q$ give the Fourier components of the potential associated with the exciton.

To construct the charge density response function $\chi(q,q^{\prime};\omega)$, we perform a real-time time-dependent density functional theory (rt-TDDFT) calculation of the valence charge density response $\rho(t)$ to the potential $V_q$ created by the core-excited state.  Optimally, one should build the perturbation $V_q$ from the full exciton wave-functions of the combined core-hole and photo-electron, as has recently been proposed \cite{Cudazzo_satellites}.  In this case, separate rt-TDDFT calculations must be performed for each BSE eigenstate $\Omega^{\lambda}$ resulting in unique spectral functions $A_{\lambda}$ for each feature in the XAS spectrum.  To simplify our present demonstration, we follow Woicik {\em et al.}~\cite{woicik_k-edges, woicik_l-edges} and use the potential of only the core-hole in the rt-TDDFT response calculation and obtain a single spectral function for all features in the XAS spectrum.  For metallic systems, we expect this to be a fair approximation, however, it could be improved upon in future work by using the full exciton potential for each BSE eigenstate.

\begin{figure}
\includegraphics[width=5.5cm,angle=270]{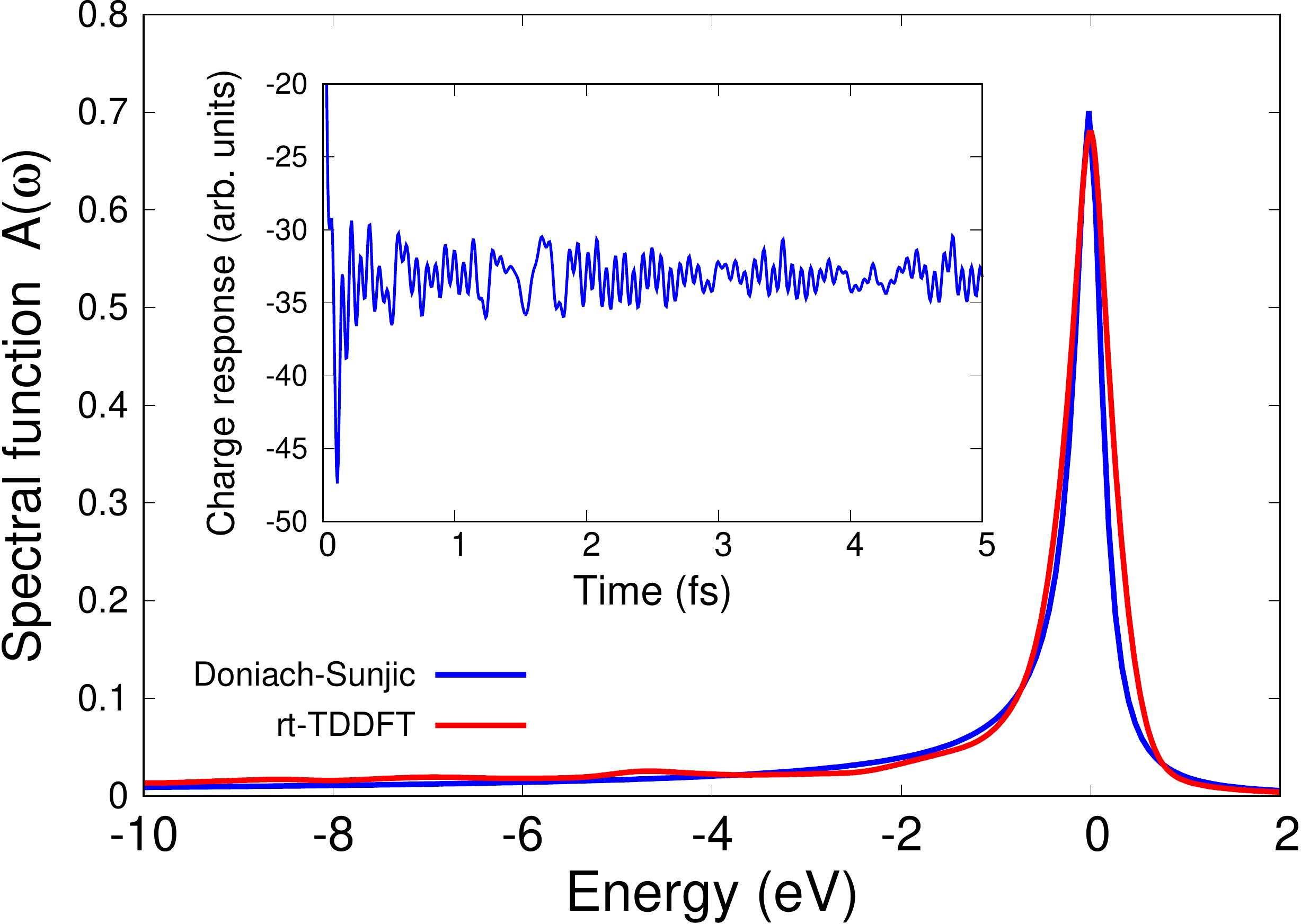}
\caption{\label{DS-XPS} Spectral function for the Fe $2p$ core-excited state in \BFA~due to secondary electronic excitations.  The red curve is the result from a rt-TDDFT calculation as described in the text.  The blue curve is a Doniach-Sunjic lineshape with an asymmetry parameter of 0.14 and a linewidth of 0.16 eV.  Inset: Real-time valence charge density response to the sudden creation of a core-hole at a Fe site.}
\end{figure}

The density response in the time domain and the resulting spectral function are shown in Fig.~\ref{DS-XPS}.  The spectral function has very little structure beyond an asymmetric tail and can be approximated well with a Doniach-Sunjic lineshape having an asymmetry parameter of 0.14. This indicates that the secondary excitations created during the x-ray absorption process consist overwhelmingly of low energy incoherent electron-hole pairs rather than distinct, localized $dd$ excitations or higher energy coherent plasmons. 

The ability to calculate effective spectral functions from first-principles with the cumulant expansion significantly improves the agreement between calculated and measured spectra without introducing any {\em ad hoc} parameters.  The technique was recently used to correctly identify excitonic, charge-transfer and shake-up features in XAS and XPS spectra at both the K- and L-edges of transition metal compounds \cite{woicik_k-edges,woicik_l-edges}.  In the next two sections, we use the cumulant spectral function to consider the impact of secondary excitations on RIXS spectra.  In particular, we provide a route to generating the indirect RIXS contributions that have previously been inaccessible to band-structure-based first-principles calculations.

\section{Incorporating secondary excitations within RIXS calculations}
\label{sect:fullrixs}

Representing the full many-body absorption and emission spectra as convolutions of the BSE spectra and spectral functions as in Eq.~\ref{convolution} offers both qualitative and quantitative advantages.  As shown above, it allows for far more accurate XAS and XES spectra of periodic systems based entirely on first-principles calculations, the accuracy of which may be systematically improved.  Here, we demonstrate the benefits for the evaluation of RIXS spectra.

Use of many-body spectral functions allows us to study the impact of secondary excitations on the direct RIXS contribution.  In this regard, the results below provide additional confirmation that the majority of the RIXS intensity in \BFA~is due to direct fluorescence effects and that the Raman-to-fluorescence crossover is a core-hole-lifetime effect within the direct contribution.  The spectral function representation also opens the possibility of constructing an effective description of the indirect contribution, which has previously been beyond the reach of first-principles calculations on periodic systems.

The basic approximation we make is to express the RIXS loss profile as a convolution of the XAS and XES spectra \cite{kas_rixs, Glover_Ge-RIXS}
~
\begin{equation}
\label{drixs-model}
    \sigma(\omega_i,\omega_o) = \int d\tilde{\omega} \frac{\mu_{a}(\tilde{\omega}) \mu_{e}(\tilde{\omega} - \omega_{l})}{(\omega_i - \tilde{\omega})^2 + (\gamma/2)^2} \, .
\end{equation}

\noindent The core-hole lifetime $(\gamma/2)^{-1}$ broadens the absorption resonance condition, allowing for the generation of intermediate-state excitons with energies within a finite range of the incident photon energy $\omega_i$.  The approximation in Eq.~\ref{drixs-model} neglects any coherence between the absorption and emission processes except that the emission intensity is evaluated at an effective emission energy $\tilde{\omega}-\omega_l$ that is shifted by the energy loss $\omega_l=\omega_i-\omega_o$ with respect to the intermediate-state exciton energy.  While the neglect of coherent contributions is not always a reliable approximation, it is often reasonable for metals.  A further approximation made in Eq.~\ref{drixs-model} is that the final-state excitonic interactions are negligible, which is also reasonable in the case of metals.

An advantage of Eq.~\eqref{drixs-model} is that it simplifies the incorporation of secondary excitations into first-principles calculations.  Specifically, taking the absorption (emission) spectra $\mu_{a}$ ($\mu_{e}$) in Eq.~\ref{drixs-model} as the convolutions in Eq.~\ref{convolution}, one can construct an effective, first-principles many-electron RIXS response for a periodic system by calculating the one- or two-particle absorption (emission) spectra $\mu_{a}^0$ ($\mu_{e}^0$), {\em e.g.} with the BSE, and then generate corresponding spectral functions $A$ as described in the previous section.  To simplify the presentation, we will assume Doniach-Sunjic lineshapes for both the absorption ($A_a$) and emission ($A_e$) spectral functions within this section.  Based on the DS fit to the rt-TDDFT spectral function shown in Fig.~\ref{DS-XPS}, we use an asymmetry parameter of 0.14 and a linewidth of 0.16 eV for the absorption spectral function.  For the emission spectral function, we reduce the asymmetry parameter to 0.07 while maintaining the linewidth of 0.16 eV, based on the favorable convolution result in Fig.~\ref{xes1}.  See Appx.~\ref{append:comp.details} for further justification of these DS lineshape parameters.

\begin{figure}
\includegraphics[width=7.5cm,angle=0]{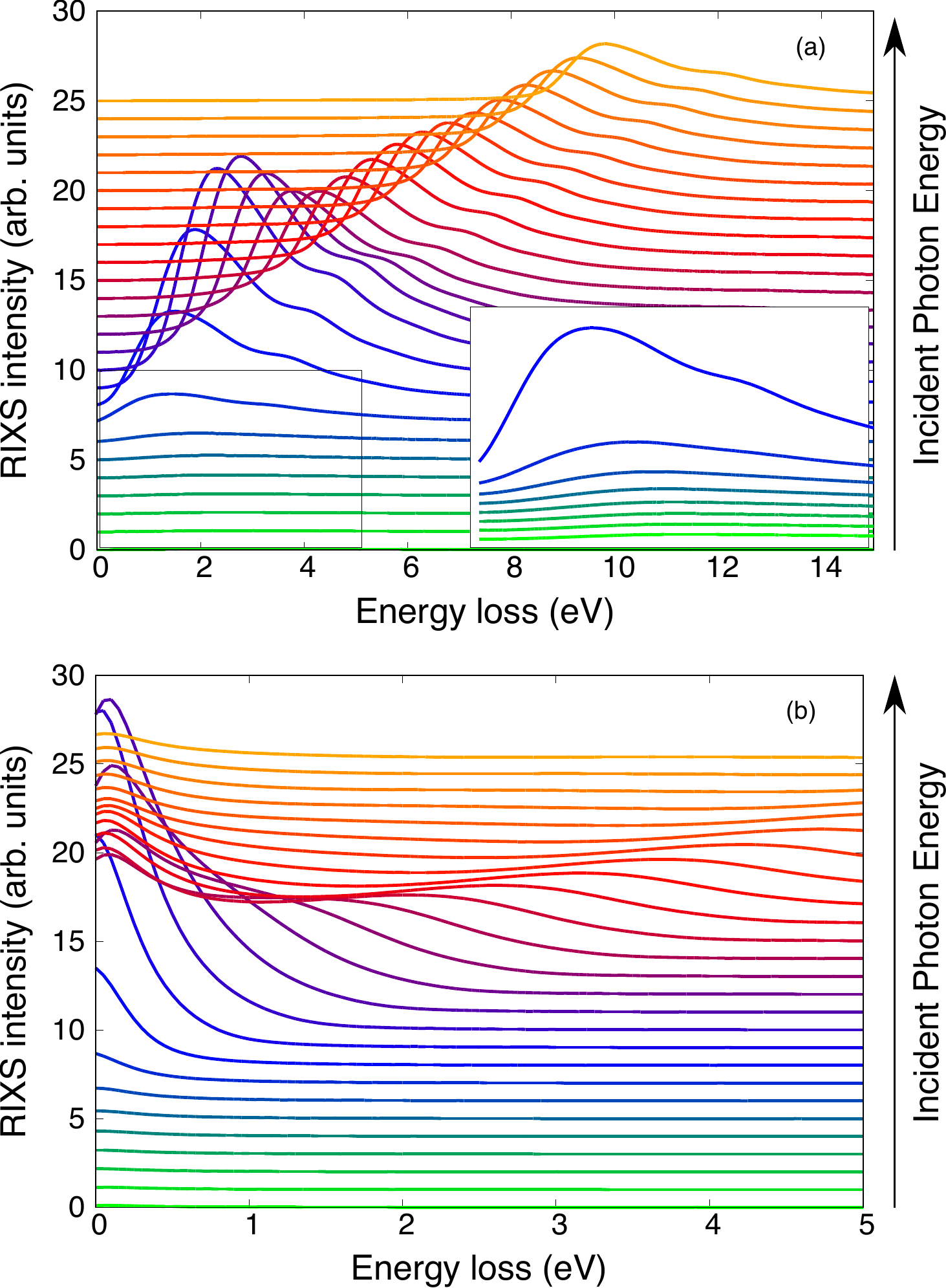}
\caption{\label{directrixs} Calculation of the direct (a) and indirect (b) RIXS contributions at the Fe L$_3$ edge of BaFe$_2$As$_2$ according to Eqs.~\ref{drixs-model} and \ref{irixs-model}, respectively.  Loss profiles for selected incident photon energies, from 702.5 eV to 715 eV in steps of 0.5 eV, are displaced vertically.  The calculations of the direct channel (a) reproduce the Raman-to-fluorescence crossover at the absorption threshold.  The inset provides an expanded view of the first 5 eV of loss for the below-threshold Raman-like region.  Within the indirect channel (b), the elastic line extends to finite energy loss related to secondary electron-hole pair excitations captured by the Doniach-Sunjic spectral function.  The unexpected fluorescence feature in (b) is discussed in the text.}
\end{figure}

The direct RIXS spectra based on Eq.~\ref{drixs-model} are shown in Fig.~\ref{directrixs}a.  The results are in good agreement with the experimental spectra and with the calculations of the full direct contribution presented in Fig.~\ref{rixs1}, supporting the appropriateness of Eq.~\ref{drixs-model} for iron-pnictides and -chalcogenides as well as more metallic systems.  The comparison confirms that the coherence between the absorption and emission processes, and the final-state excitonic effects, makes only a minor impact on the intensities of spectral features and does not noticeably affect qualitative trends.  In particular, these simplified calculations succeed in reproducing the Raman-to-fluorescence crossover as the incident photon energy traverses the absorption edge.

%\textcolor{blue}{New paragraph and figure to compare direct RIXS results with and without spectral functions.  How do the secondary excitations affect the spectra?  Particularly, do they contribute a linear background at low energy loss above the XAS threshold?  What about the decay profile beyond the peak of the RIXS loss?}

In \BFA, the x-ray absorption process evidently generates secondary excitations to a greater extent than the emission process.  Therefore, a further approximation we make for the indirect RIXS contribution is that the energy loss originates from those secondary excitations occurring during the absorption step.  To make this more evident, we rewrite the full absorption spectrum $\mu_a$ as an explicit product over the bare BSE spectrum $\mu_{a}^{0}$ at an energy $\tilde{\omega}$ corresponding to the primary exciton and the spectral function $A_{a}$ evaluated at the energy loss $\omega_{l}$ such that the intermediate-state exciton energy is given by $\tilde{\omega}+\omega_l$.  Furthermore, we replace the full XES spectrum $\mu_e$ with the bare BSE spectrum $\mu_{e}^{0}$.  Since for the indirect contribution, the state involved in the emission process is the same excitonic state created during the absorption step, we replace the emission factor $\mu_{e}^{0}(\tilde{\omega})$ with a second absorption factor $\mu_{a}^{0}(\tilde{\omega})$.  The indirect RIXS contribution reduces to
~
\begin{equation}
\label{irixs-model}
    \sigma(\omega_i,\omega_o) = \int d\tilde{\omega} \frac{ \left [ \mu_{a}^{0}(\tilde{\omega}) \right ]^2 A_{a}(\omega_{l})}{(\omega_i - \tilde{\omega} - \omega_{l})^2 + (\gamma/2)^2} \, .
\end{equation}

\noindent The resonant denominator selects for an incident photon energy $\omega_i \approx \tilde{\omega}+\omega_{l}$ while the outgoing photon has energy $\omega_o = \tilde{\omega}$.

Results for the indirect RIXS contribution are presented in Fig.~\ref{directrixs}b.  As expected given the fairly featureless structure of the Doniach-Sunjic spectral function, the indirect RIXS signal consists mainly of a Raman-like quasi-elastic line with a tail that extends to finite energy loss.  The intensity of the quasi-elastic feature varies with the absorption profile.  An unexpected fluorescence-like feature is also clearly present.  This behavior originates in the highly non-uniform absorption profile at the Fe L-edge.  Due to the much greater absorption intensity at the main peak compared to the post-peak region, secondary excitations associated with excitonic states at the main peak make sizeable contributions to the loss profile.  As the incident photon energy is tuned further above the main edge, the energy of the secondary excitations associated with the main edge increase accordingly, causing an apparent fluorescence-like behavior for the loss peak.  This effect has potentially been observed experimentally in LAO/STO superlattices \cite{zhou_localized_2011}.

\begin{figure}
\centering
\includegraphics[width=7.5cm,angle=0]{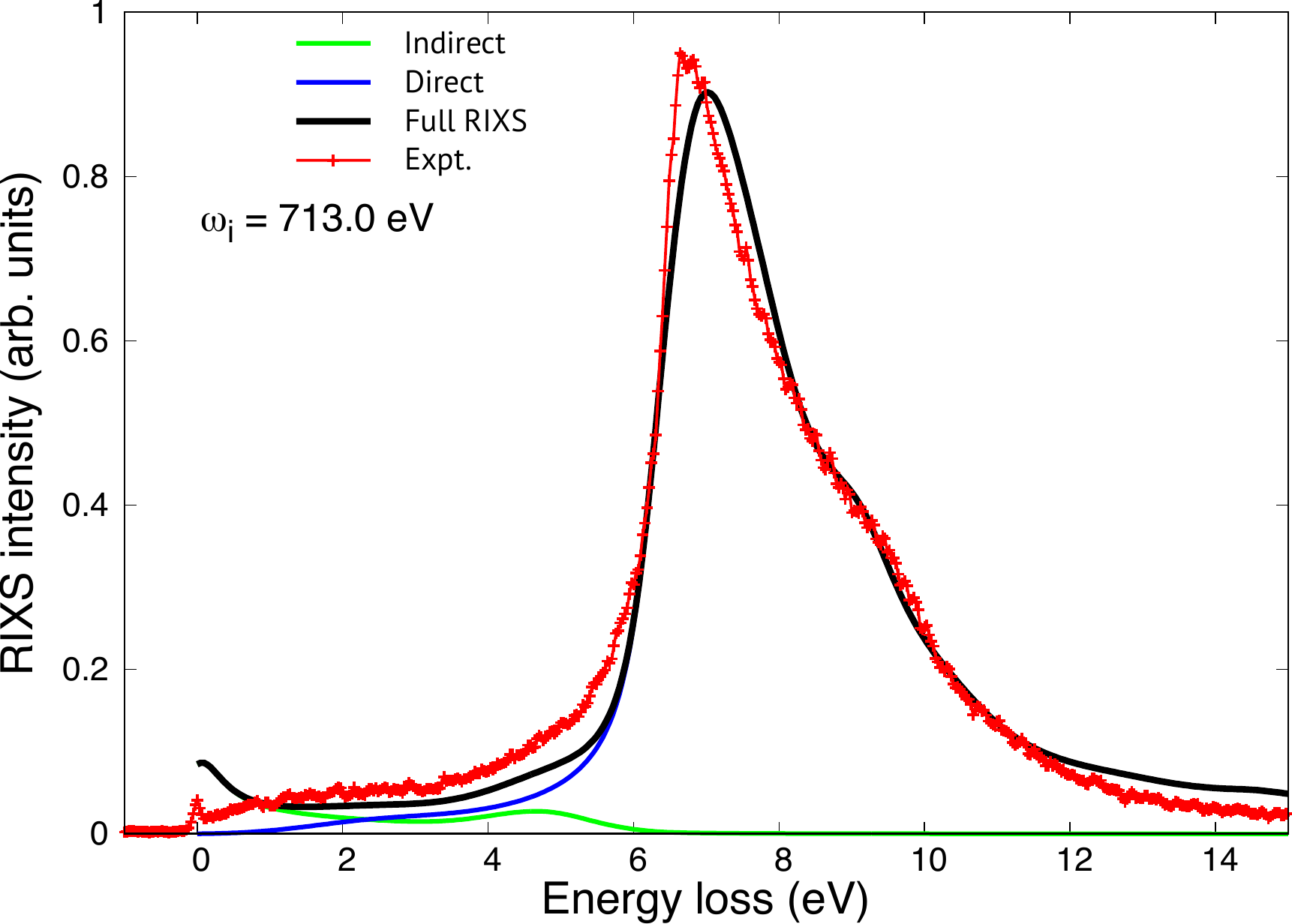}
\caption{\label{rixscuts} Measured and calculated full RIXS profile for BaFe$_2$As$_2$ for an incident photon energy of 713.0 eV.  Calculated indirect (green) and direct (blue) contributions are shown in addition to the full RIXS signal (black) with comparison to the experimental result (red symbols).  The energy scale of the calculated RIXS profiles have been contracted by 10$\%$ to correct for known deficiencies of the LDA treatment of correlations in \BFA~\cite{Tomczak_bafe2as2}.}
\end{figure}

Figure \ref{rixscuts} shows that the calculations based on Eqs.~\ref{drixs-model} and \ref{irixs-model} compare very well to the experimental results after a 10$\%$ contraction of the energy scale for the calculations.  This energy scale contraction is necessary both to correct the LDA bandwidth of \BFA \cite{Tomczak_bafe2as2} and because final-state excitonic binding has been neglected within the present approximations, though this is likely a smaller effect.  This comparison further demonstrates that the large majority of the RIXS signal occurs through the direct channel, and additionally validates the approximations made in Eqs.~\ref{drixs-model} and \ref{irixs-model}.  Interestingly, the experimental result shows additional intensity compared to the direct contribution around 5 eV.  Here, the fluorescence-like contribution to the indirect signal augments the direct contribution, noticeably improving the agreement with experiment.

\section{Observation of $\boldsymbol{dd}$ excitations in ${\rm \mathbf{BaFe_2As_2}}$} \label{sect:model}
\begin{figure}
\centering
\includegraphics[scale =0.55]{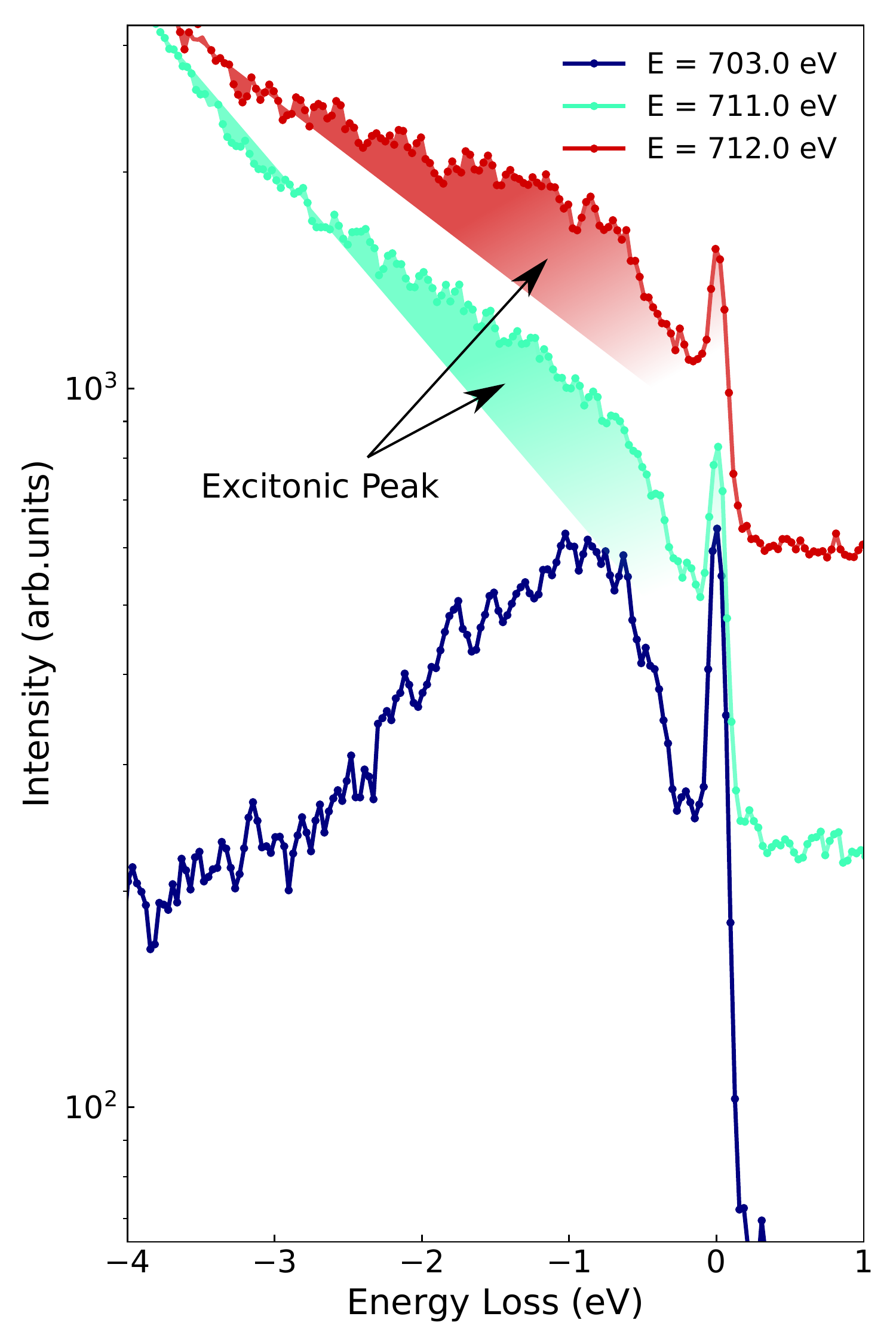}
\caption{\label{fig7} (a) Comparison of the excitonic peak at $\omega_l \approx$ 1.0 eV from measured RIXS loss profiles of \BFA~for incident photon energies below and above the absorption edge. The spectra have been offset and normalized for better visualization. The intensity scale is logarithmic.}
\end{figure}

A hallmark of strongly correlated electron systems is the loss of quasiparticle coherence and the emergence of Hubbard sidebands.  In terms of photoemission, the sudden removal of an electron is ubiquitously accompanied by the generation of secondary excitations that appear as satellite features to the main spectral line with intensities determined by the quasiparticle renormalization factor.  In this respect, it was surprising that early RIXS studies on Fe pnictides did not show clear \textit{dd} excitations \cite{yang_evidence_2009}. However, more recent work displayed orbital excitations \cite{nomura_resonant_2016}. The lack of obvious pure indirect Raman features in the RIXS profiles of Fe pnictides is seemingly at odds with their correlated nature.  Such features would be expected to appear at low energy loss, within the localized iron $d$ bands.  Close inspection of our experimental results reveals the persistence of a faint peak between about 1-2 eV at high incident energy, as shown in Fig.~\ref{fig7}. The evidence that this peak changes in width and intensity but not in energy as a function of incident energy demonstrates that it is a coherent excitation of excitonic nature.  This excitonic feature has not been observed or discussed in the literature, likely because at low incident energy the fluorescence and exciton peaks overlap, precluding the detection of the latter in Fe pnictides so far.  Tuning to high incident energy (well above the threshold) shifts the fluorescence signal away, uncovering the excitonic peak.

\begin{figure}
\centering
\includegraphics[width=6cm,angle=270]{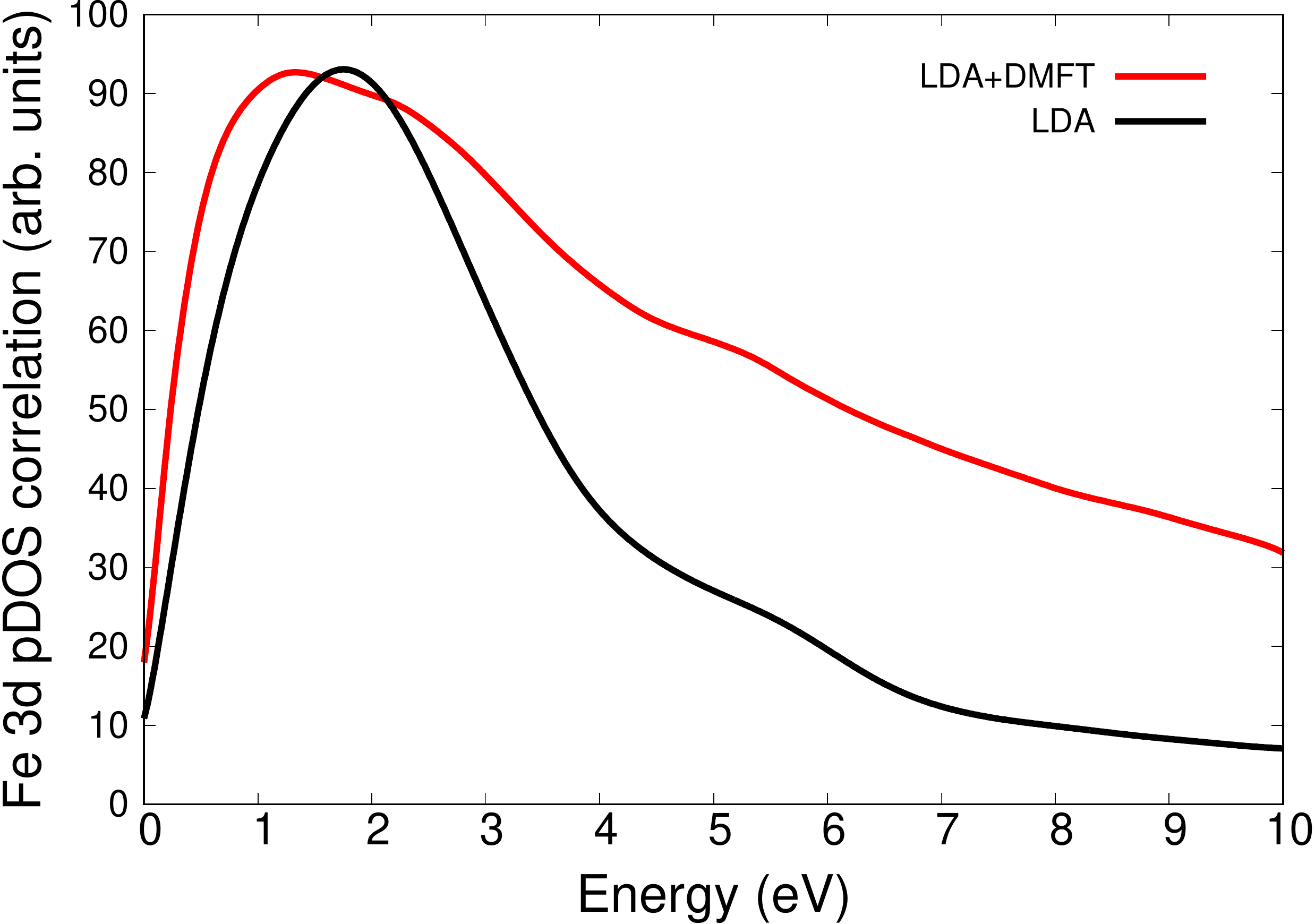}
\caption{\label{autocorr} Correlation function of the occupied and unoccupied Fe 3$d$ pDOS generated at the level of LDA (black) and LDA+DMFT (red).}
\end{figure}

If the excitonic feature we observe in Fig.~\ref{fig7} is indeed due to a $dd$ excitation on the Fe site then the energy position and structure of this peak should correspond roughly to the correlation function between the occupied and unoccupied pDOS of the Fe 3$d$ orbitals.  To confirm this, we evaluate this correlation function as
~
\begin{equation}
    \sum_{i} \int d\omega^{\prime} \rho_{occ}^{i}(\omega^{\prime}) \rho_{unocc}^{i}(\omega^{\prime}+\omega)
\end{equation}

\noindent where $i$ indicates a specific Fe 3$d$ orbital and $\rho^{i}$ is the density of states for that orbital.  The resulting correlation functions for the LDA and LDA+DMFT densities of states shown in Fig.~\ref{pDOS} are given in Fig.~\ref{autocorr}.  Whereas the LDA pDOS correlation function peaks narrowly around 1.8 eV, the LDA+DMFT pDOS correlation function presents a broader double-peak structure spanning from about 0.9-2.5 eV.  The LDA+DMFT correlation function closely resembles the excitonic peak indicated in Fig.~\ref{fig7}, supporting the assignment of this feature as an indirect channel $dd$ excitation.

Despite the clear connection between the pDOS correlation function and the 1-2 eV excitonic feature, our calculations of the indirect RIXS channel (Fig.~\ref{directrixs}b) do not show any distinct $dd$ excitation in this energy range even when the incident photon energy is tuned well above threshold.  Within our computational scheme, indirect effects are captured by the excitonic spectral function, which we have generated through a rt-TDDFT calculation using the generalized gradient approximation (GGA) to the exchange-correlation functional.  This yielded a profile very similar to a Doniach-Sunjic lineshape, which, after convolution with the BSE spectra, produced quite accurate XAS and XES spectra compared to the experimental results.  The DS lineshape also provides a very good fit to the experimental x-ray photoemission spectrum of \BFA.  However, it is conceivable that our TDDFT calculation might fail to capture certain on-site correlated responses for \BFA~that would contribute a distinct $dd$ excitonic feature to the otherwise featureless DS profile obtained for the spectral function.  This could also be a source of the small quantitative discrepancies in the indirect RIXS contribution shown in Fig.~\ref{rixscuts}.  Therefore, it is instructive to consider the RIXS spectra resulting from a spectral function that is more structured, having a distinct $dd$ excitonic satellite.

\begin{figure*}
\centering
 \begin{subfigure}
  \centering
  \includegraphics[scale=0.2675,angle=270]{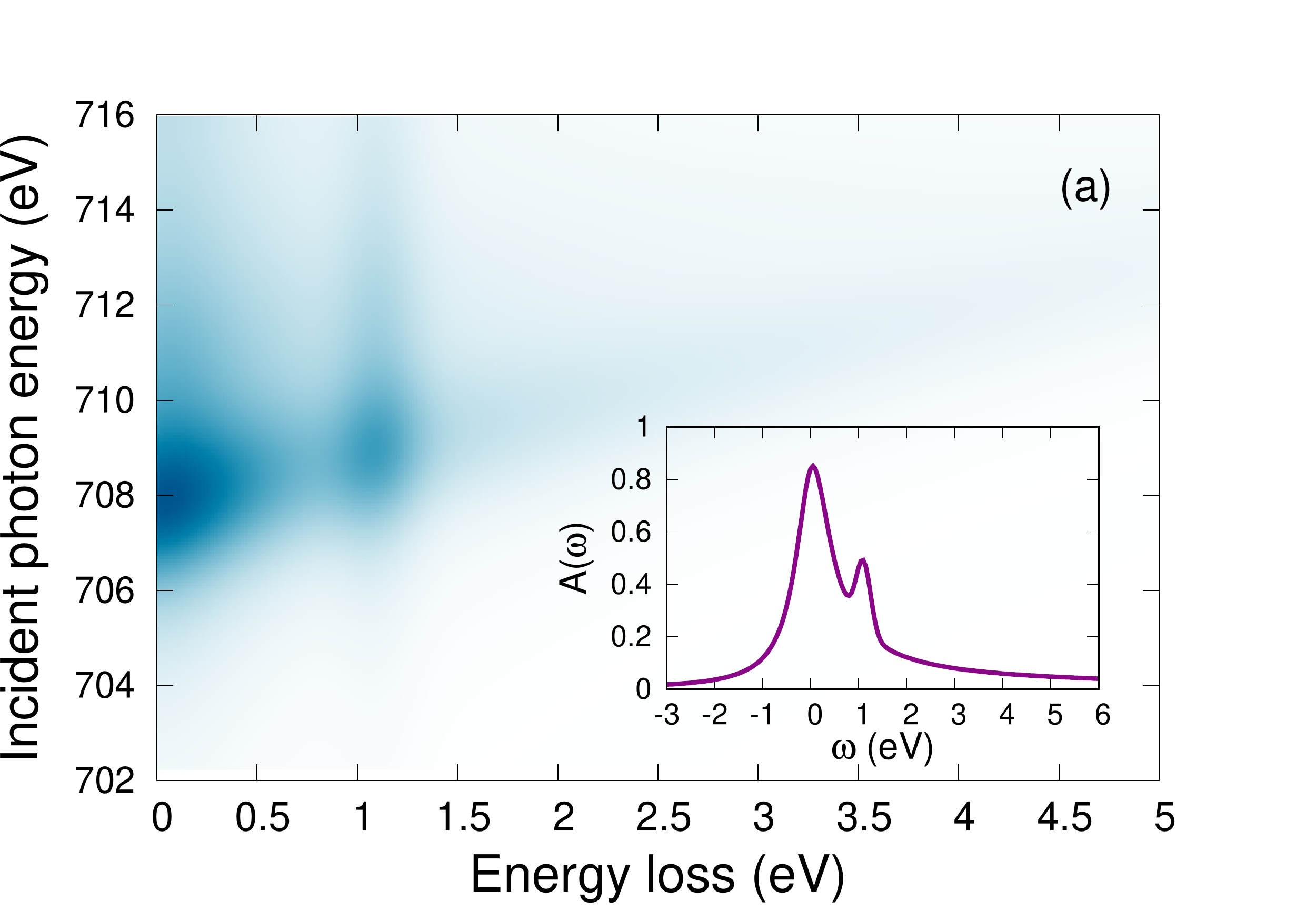}
 \end{subfigure}
\hspace{-1.6cm}
 \begin{subfigure}
  \centering
  \includegraphics[scale=0.2675,angle=270]{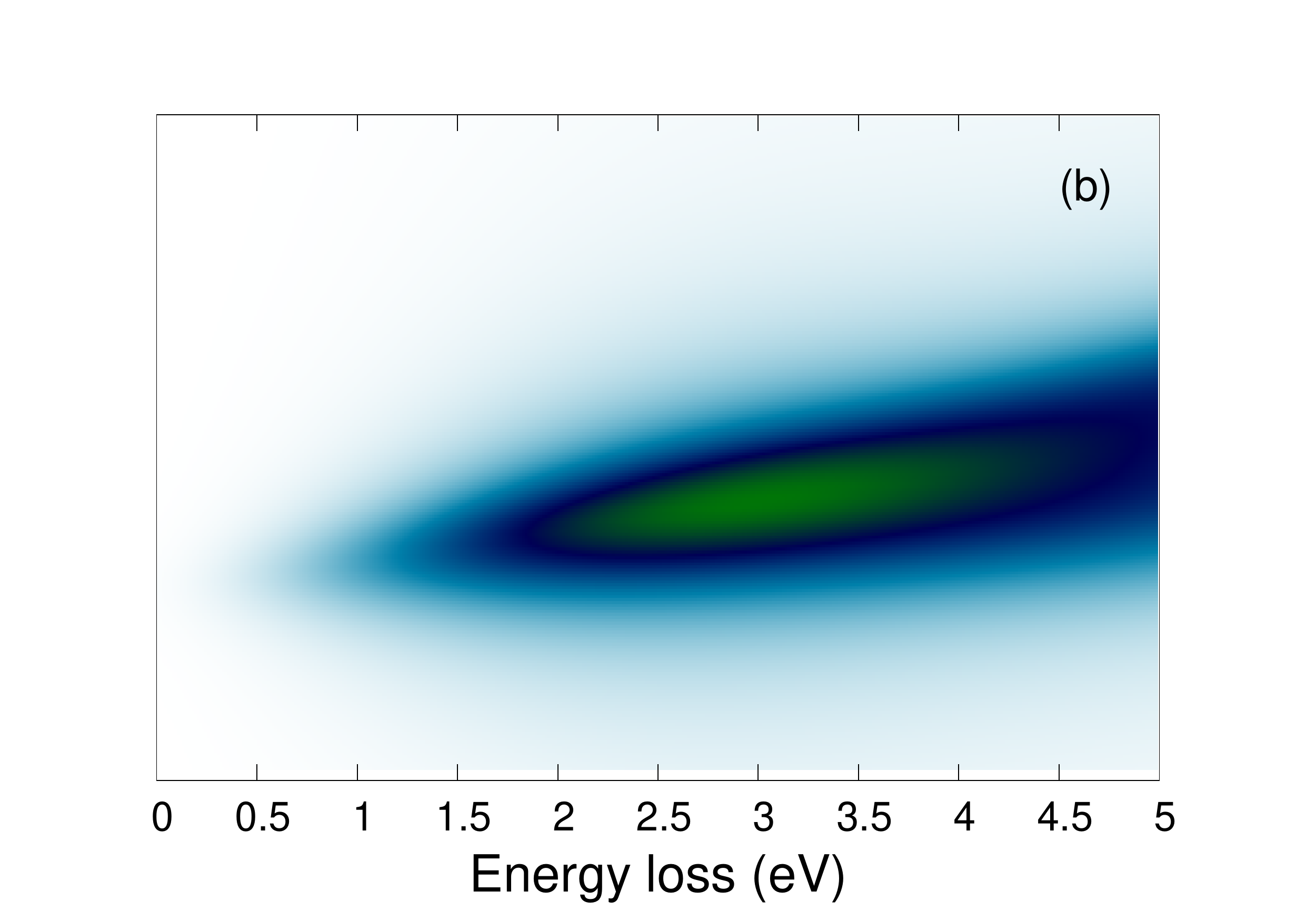}
 \end{subfigure}
 \hspace{-1.6cm}
 \begin{subfigure}
  \centering
  \includegraphics[scale=0.2675,angle=270]{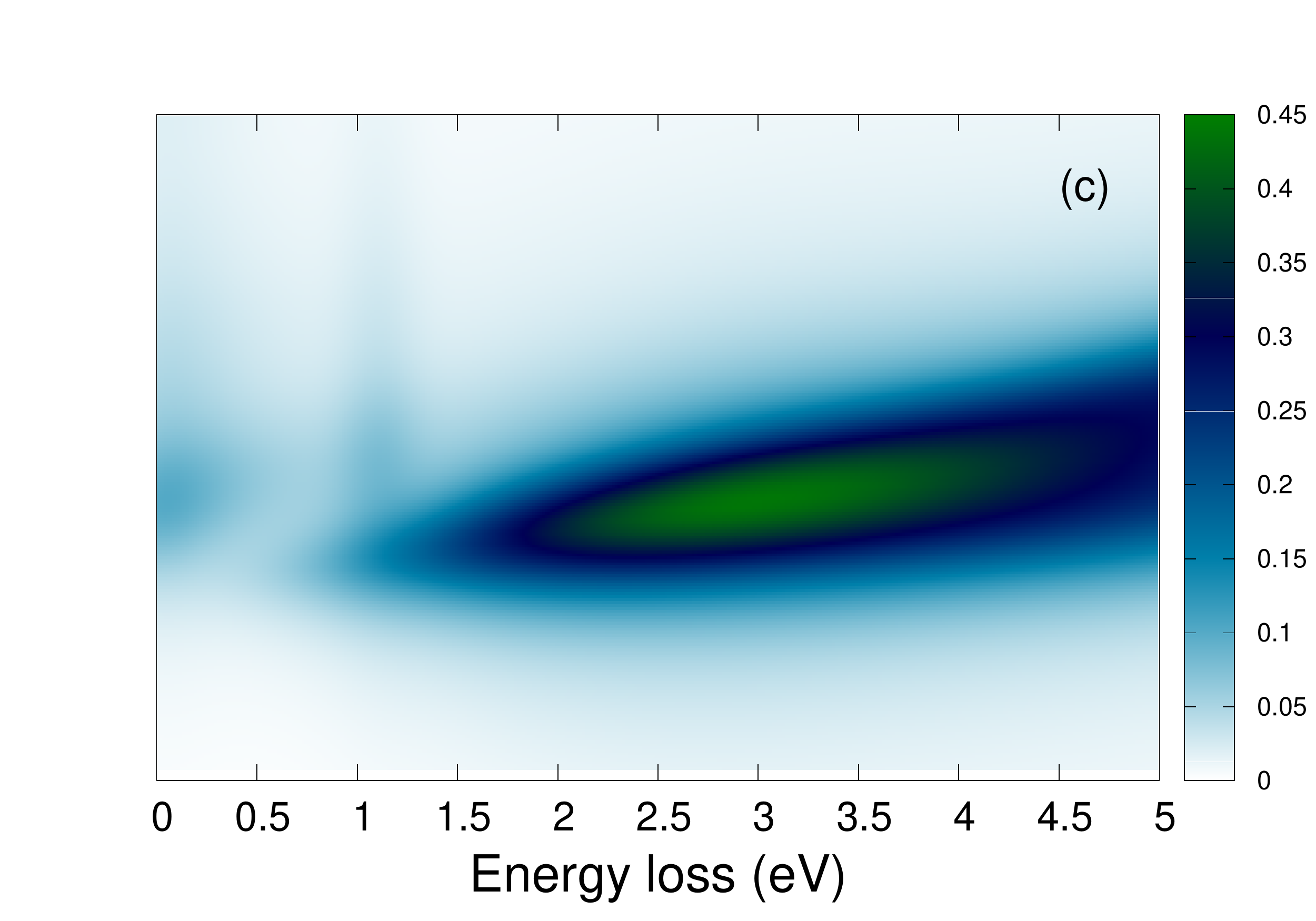}
 \end{subfigure}
\caption{\label{rixs-dd} Calculated RIXS maps for \BFA~using the model spectral function shown in the inset of (a) in conjunction with the approximations made in Eqs.~\ref{drixs-model} and \ref{irixs-model}.  The panels show the indirect contribution (a), direct contribution (b), and the full RIXS signal (c). The intensity of the indirect contribution in panel (a) has been multiplied by 2 for visual clarity. The model spectral function consists of a Doniach-Sunjic profile with the addition of a satellite peak at 1.1 eV representing a local $dd$ excitation.} %\textcolor{green}{Replace with high-resolution versions before submitting.}}
\end{figure*}

We present this case in Fig.~\ref{rixs-dd} by combining the bare BSE absorption and emission spectra ($\mu_a^0$ and $\mu_e^0$), calculated specifically for BaFe$_2$As$_2$, with an artificially constructed spectral function consisting of a dominant DS profile with an additional satellite peak split off by 1.1 eV (see the inset of Fig.~\ref{rixs-dd}a).  The indirect contribution (Fig.~\ref{rixs-dd}a) consists of a true elastic feature at zero energy loss  and its fluorescent extension as in Fig.~\ref{directrixs}b, and a weaker Raman feature at 1.1 eV loss associated with the $dd$ satellite of the spectral function.  We note that the maximum in intensity of the 1.1 eV Raman loss feature occurs for an incident energy 1.1 eV above the peak of the elastic line ({\em i.e.}~it is fluorescence shifted).  This model result resembles the measured spectra in the low loss region well above threshold and further confirms that we observed a $dd$ excitation that is not generated by the TDDFT calculation using a GGA exchange-correlation functional.

For completeness, we also show the direct RIXS contribution (Fig.~\ref{rixs-dd}b) for this model spectral function.  The direct RIXS profile is again dominated by fluorescence above threshold.  Below threshold, a Raman-like peak is still present, but too weak to be observed in the color map.  The indirect and direct contributions are combined in Fig.~\ref{rixs-dd}c.

%%% The high-res versions are about 9M together and cause compilation to be very slow so I inserted a temporary low-res version.  You can switch them if you want to generate a pdf.

\section{Discussion and Conclusions} \label{conclusions}
The incident photon energy dependence of the RIXS response of \BFA~exhibits a clear crossover from Raman-like behavior below the XAS threshold to fluorescence-like above the XAS threshold.  Our analysis of the spectral features, paired with first-principles calculations, indicate that in both regions the loss spectra are dominated by the direct RIXS response. The Raman-like behavior in the direct RIXS channel, seen below the XAS threshold, can be understood as a consequence of the core-hole lifetime broadening.  By pairing experimental and computational analysis, we uncovered the persistence of a $dd$ exciton in the indirect channel, previously undetected in the Fe pnictides, indicating at least a partial localization of the electronic states of \BFA.

We have demonstrated the ability to calculate RIXS spectra from first-principles with full momentum dependence for metallic systems.  These calculations, performed within the framework of the Bethe-Salpeter equations, are limited to a two-particle description of excited states, which is often insufficient for metallic and strongly correlated materials.  However, we have shown how to overcome this limitation by convolving the BSE spectra with effective many-body spectral functions.  This greatly improved the agreement between calculated and measured x-ray absorption and emission spectra, and allowed, for the first-time, {\em ab initio} evaluation of the indirect RIXS response accounting for the full periodicity of the system.

In the present work, we have calculated the effective spectral functions using rt-TDDFT with a GGA exchange-correlation functional, which failed to generate the $dd$ excitation observed experimentally.  In the future, we envisage that the spectral functions, which account for the generation of secondary excitations during the dynamic screening response of the system, could be calculated at a more rigorous level of theory for correlated electron materials.  Possibilities include many electron wavefunction methods applied to a local cluster, extensions of dynamical mean field theory, or explicit double exciton representations.

We believe that our work can be used as the basis for more detailed studies of the RIXS response in correlated metals that can lead to refined methodologies for extracting materials characteristics and intrinsic parameters from the energy dependence of the RIXS response. Additionally the ability of DFT and \textit{ab initio} calculations to deal with multi-orbital systems will allow the extension of these calculations to complex systems of enhanced spin-orbital complexity such as cobaltates and manganites.

\begin{acknowledgments}
K.G.  was supported by the U.S. Department of Energy, Office of Science, Basic Energy Sciences as part of the Computational Materials Science Program. J.P. and T.S. acknowledge financial support through the Dysenos AG by Kabelwerke Brugg AG Holding, Fachhochschule Nordwestschweiz, and the Paul Scherrer Institut. J. P. acknowledges financial support by the Swiss National Science Foundation Early Postdoc. Mobility fellowship  Project  No. P2FRP2\_171824 and P400P2\_180744. The synchrotron radiation experiments have been performed at the ADRESS beamline of the Swiss Light Source at the Paul Scherrer Institut. Part of this research has been funded by the Swiss National Science Foundation through the D-A-CH program (SNSF Research Grant No. 200021L 141325). Work in Japan was supported by Grant-in-Aids for Scientific Research (KAKENHI) from Japan Society for the Promotion of Science (JSPS), and by the `Quantum Liquid Crystals' Grant-in-Aid  for Scientific Research on Innovative Areas from the Ministry of Education, Culture, Sports, Science and Technology (MEXT) of Japan.

K.G. and J.P. contributed equally to this work.  K.G. would like to thank Joe Woicik and Ignace Jarrige for valuable discussions.
\end{acknowledgments}

\appendix

\section{Experimental methods} \label{exp}
Single crystals of \BFA~were grown by stoichiometric melt as described in \cite{kasahara_evolution_2010} and characterized either with resistivity or magnetization measurements \cite{kasahara_evolution_2010}.  Fe-L$_{2,3}$ XAS and RIXS experiments were performed at the ADRESS beamline of the Swiss Light Source, Paul Scherrer Institut, Villigen PSI, Switzerland \cite{strocov_high-resolution_2010,ghiringhelli_saxes_2006,schmitt_high-resolution_2013}. XAS was measured in both total fluorescence yield (TFY) and total electron yield (TEY) without observing significant differences. XAS and RIXS spectra were recorded at 15$^{\circ}$ incidence angle relative to the sample surface. The RIXS spectrometer was set to a scattering angle of 130 degrees. The total energy resolution was measured employing the elastic scattering of carbon-filled acrylic tape and was around 110 meV. The samples were mounted for XAS and RIXS experiments with the \textit{ab} plane perpendicular to the scattering plane and the \textit{c} axis lying in it and post-cleaved \textit{in situ} at a pressure better than 2.0 $\times$ 10$^{-10}$ mbar. We measured at (0.31, 0.31) using the orthorhombic unfolded crystallographic notation \cite{park_symmetry_2010}. All the measurements were carried out at 14~K by cooling the manipulator with liquid helium. 

\BFA~has the tetragonal lattice structure (space group I4/mmm) with 2 Fe unit cell, and the relaxed parameters are \textit{a}$=$3.868 \AA~and \textit{c}$=$12.378 \AA.

\section{Computational details}  \label{append:comp.details}

All Bethe-Salpeter equation (BSE) spectral calculations were performed with the OCEAN code \cite{gilmore_efficient_2015} based on a density functional theory (DFT) electronic structure obtained with Quantum Espresso \cite{Gianozzi_QE-2009, Gianozzi_QE-2017}, which uses pseudopotentials and a planewave basis.  The DFT calculations employed the local density approximation to the exchange-correlation functional and used norm conserving pseudopotentials.  For Fe, we constructed a pseudopotential with semi-core states in valence.

For the x-ray absorption and emission calculations, a ground-state electronic charge density was obtained from a self-consistent-field (SCF) calculation with a 3x3x2 k-point sampling.  To evaluate the BSE spectrum, we constructed a basis set of Bloch state by performing a non self-consistent-field (NSCF) DFT calculation with 7x7x5 k-point sampling and 160 unoccupied bands for the absorption calculation.  Both the SCF and NSCF calculations used a planewave cutoff energy of 120 Ry for the wavefunctions.  Due to the need to treat the final-state valence excitonic states, the RIXS calculations required a much denser k-point sampling of 16x16x10.

The final-state density of states (DOS) calculations were performed with the Quantum Espresso code taking a 2x2x1 supercell of the conventional unit cell.  For these calculations, we used ultrasoft pseudopotentials and the generalized gradient approximation to the exchange-correlation functional.  The k-point samplings were 2x2x3 for the SCF calculation and 5x5x8 for the NSCF calculation.  The energy cutoffs were 40 Ry for the wavefunctions and 320 Ry for the charge density.  For comparison to the XAS, the pseudopotential for one Fe site was replaced with an alternate pseudopotential containing a hole in the 2$p$ shell while an extra electron was placed at the bottom of the conduction band to simulate the XAS final state.  The ground-state DOS, for comparison to the XES, was calculated analogously, but without the use of a supercell or core-hole pseudopotential and extra electron.

The real-time TDDFT calculation, used to generate the valence charge density response to the sudden creation of an excitonic state, was performed with a modification \cite{takimoto_rttddft} of the SIESTA DFT code \cite{Soler_Siesta-2002}.  The LDA-DFT based dynamical mean field theory calculations were performed with the Comscope code ComDMFT \cite{Choi_ComDMFT-2019}, which solves the impurity problem using a continuous-time quantum Monte Carlo algorithm.  

As we demonstrated in Sec.~\ref{sect:mnd}, the many-body spectral function associated with the sudden creation of a core-level exciton in \BFA~can be approximated very well by a Doniach-Sunjic lineshape, the expression for which is
~
\begin{equation}
    DS(\omega) = \frac{\cos{\frac{\pi\alpha}{2}+(1-\alpha)\arctan(\frac{\omega-\omega_0}{\sigma})}}{\left ( \sigma^2 + (\omega-\omega_0)^2 \right )^{(1-\alpha)/2}} \, .
\end{equation}

\noindent The key parameter is the dimensionless quantity $\alpha$, which characterizes the asymmetry of the lineshape.  The other parameters are $\omega_0$, which indicates the peak position, and $\sigma$ that gives the symmetric contribution to the linewidth.  Based on the spectral function obtained from the rt-TDDFT response, we used an asymmetry value of $\alpha=0.14$ and a linewidth of $\sigma$=0.16 eV for x-ray absorption.  By convolving a DS lineshape with the BSE calculated x-ray emission spectrum and comparing to the experimental result, we obtain an asymmetry value of $\alpha=0.07$ for the XES process and the same linewidth of $\sigma$=0.16 eV.

%Previous experimental analysis of the x-ray photoemission of \BFA~\cite{de_jong_high-resolution_2009} found $\alpha=0.44$, whereas our rt-TDDFT calculation suggests a smaller value (less asymmetry) of $\alpha=0.14$ for x-ray absorption.  This is not particularly surprising given that for XAS the excited electron partially screens and counteracts the positive core-hole.  By convolving a DS lineshape with the BSE calculated x-ray emission spectrum and comparing to the experimental result, we obtain an asymmetry value of $\alpha=0.07$ for the XES process.

\section{Lifetime effect}  \label{append:lifetime}

\begin{figure}
\centering
\includegraphics[scale=0.325,angle=270]{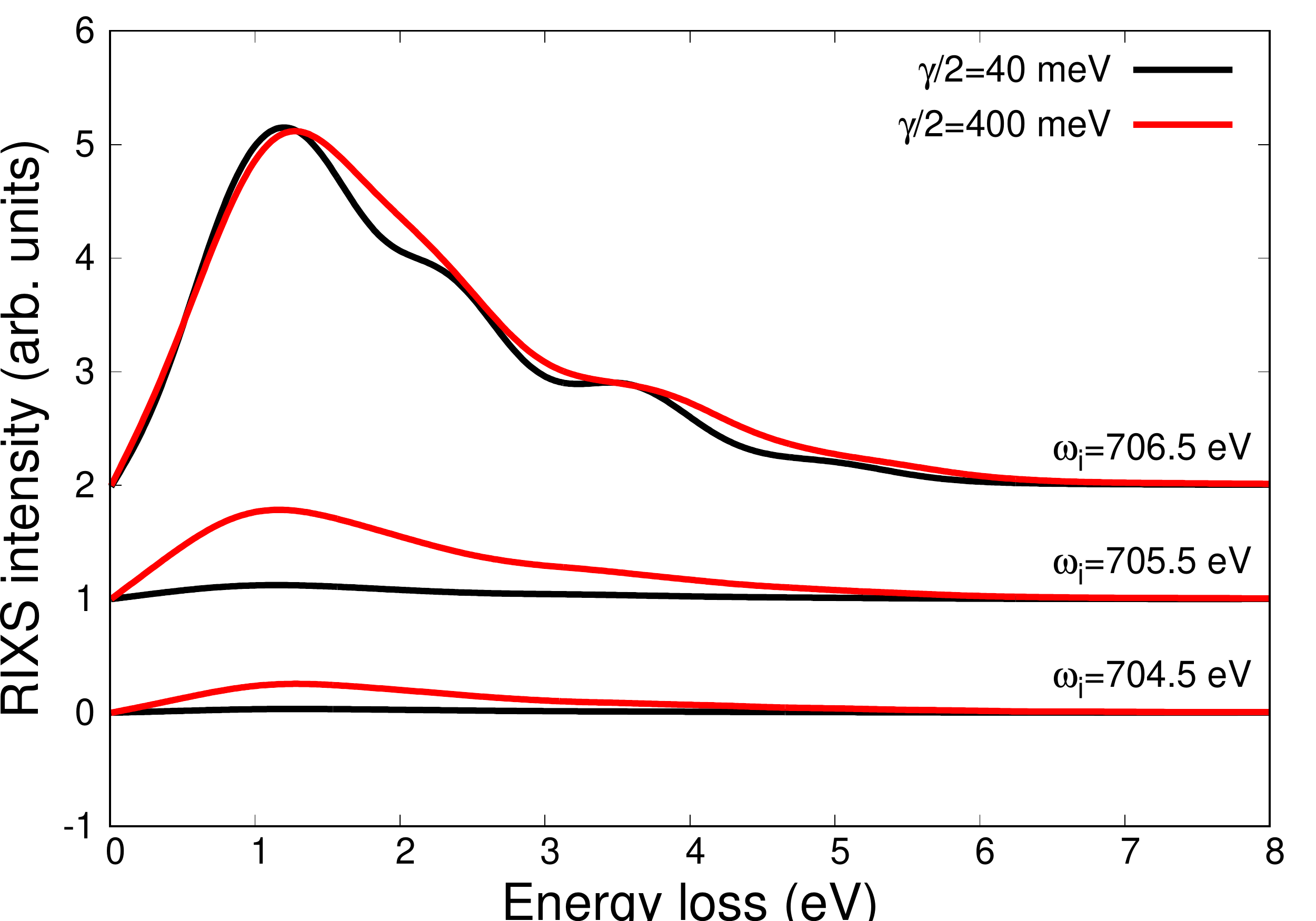}
\caption{\label{fig:broadening} Bethe-Salpeter equation calculation of the direct RIXS response of \BFA~at and below the Fe L$_3$ XAS threshold for core-level broadenings of 40 meV and 400 meV.  The longer core-hole lifetime (black) attenuates the RIXS signal below threshold more rapidly.}
\end{figure}

To further demonstrate that the Raman-like behavior of the RIXS spectra of \BFA~below the XAS threshold originates from direct RIXS fluorescence processes that are enabled by the finite core-hole lifetime, we calculate the direct RIXS contribution for two values of the core-hole lifetime.  Figure \ref{fig:broadening} shows the calculated RIXS loss profiles below the XAS threshold for two values of the core-hole lifetime broadening.  This figure verifies that the below threshold signal persists to much lower incident photon energy values for the larger core-level broadening.  This can also be understood from the denominator of the Kramers-Heisenberg equation or from our approximation in Eq.~\ref{drixs-model} which show that the signal intensity varies with the detuning $\Delta$ as $1/(\Delta^2 + (\gamma/2)^2)$.

%%% The high-res version (lifetime.eps) is about 9M and causes compilation to be very slow so I inserted a temporary low-res version.  You can switch them if you want to generate a pdf.

\end{document}